\documentclass[%
 reprint,
 amsmath,amssymb,
 aps,
pra,
floatfix,
]{revtex4-1}

\usepackage{graphicx}% Include figure files
\usepackage{dcolumn}% Align table columns on decimal point
\usepackage{hyperref}% add hypertext capabilities
\hypersetup{
anchorcolor=blue,
colorlinks=true,
citecolor=blue,
urlcolor=blue,
linkcolor=blue}

\usepackage{braket}
\usepackage{commath}
\usepackage{color}
\usepackage[caption=false]{subfig}
\usepackage{dsfont}
\usepackage{float}
\usepackage{textcomp}
\usepackage{gensymb}
\usepackage{siunitx}
\usepackage{afterpage}
\usepackage{placeins}
\usepackage{ragged2e}
\newcommand{\adagger}{\ensuremath{\hat A^\dagger}}
\newcommand{\asmalldag}{\hat a^\dagger}
\newcommand{\asmallhat}{\hat a}
\newcommand{\imi}{\text{i}}
\newcommand{\sigmaplus}{\hat \sigma_+}
\newcommand{\sigmaminus}{\hat \sigma_-}
\newcommand{\aout}{\ensuremath{\hat a_\text{out}}}

\newcommand{\ain}{\ensuremath{\hat a_\text{in}}}
\newcommand{\adagin}{\ensuremath{\hat a^\dagger_\text{in}}}

\newcommand{\dt}{\ensuremath{\dif t}}

\newcommand{\expup}[1]{\ensuremath{e^{#1}}}

\newcommand{\xhattheta}{\ensuremath{ \hat X_\theta}}
\newcommand{\xtheta}{\ensuremath{ x_\theta}}

\newcommand{\ketbra}[2]{\ensuremath{\ket{#1}\hspace{-1.5pt}\bra{#2}}}

\newcommand{\tr}{\ensuremath{\mathrm{Tr}}}
\renewcommand{\neg}{\mathsf{W}}

\newcommand{\gamman}{\Gamma_\text{n}}
 \newcommand{\gammaphi}{\Gamma_\phi}
 \newcommand{\ahat}{\hat A}
 \newcommand{\ninc}{n_\text{inc}}
 \newcommand{\ncoh}{n_\text{coh}}
 \renewcommand{\ss}[1]{\braket{#1}_\text{ss}}
\newcommand{\rhoc}{\hat\rho_\text{c}}

\begin{document}

%\preprint{APS/123-QED}

\title{Numerical study of Wigner negativity in one-dimensional steady-state resonance fluorescence }

\author{Ingrid Strandberg}
\author{Yong Lu}
\author{Fernando Quijandr\'{\i}a}
\author{G\"oran Johansson}
\affiliation{Microtechnology and Nanoscience, MC2, Chalmers University of Technology, SE-412 96
G\"oteborg, Sweden}

\date{\today}

% Moves in abstracts 
% Move 1: Background/Introduction/Situation
% Move 2: Present research/purpose
% Move 3: Methods/materials/subjects/procedures
% Move 4: Results/findings
% Move 5: Discussion/ conclusion/ implications/recommendations

%As an indicator of nonclassicality 
%We use negativity of the Wigner function as a measure of nonclassicality, since such nonclassical states can be used for quantum information processing applications, including quantum computing. 
%By solving a stochastic master equation, the state of the two-level atom is conditioned on homodyne measurements of the resonance fluorescence. From the measurement statistics we reconstruct the quantum state of the emitted field. 

\begin{abstract}

In a numerical study, we investigate the steady-state generation of nonclassical states of light from a coherently driven two-level atom in a one-dimensional waveguide. Specifically, we look for states with a negative Wigner function, since such nonclassical states are a resource for quantum information processing applications, including quantum computing. We find that a waveguide terminated by a mirror at the position of the atom can provide Wigner-negative states, while an infinite waveguide yields strictly positive Wigner functions. Moreover, our investigation reveals a connection between the purity of a quantum state and its Wigner negativity. We also analyze the effects of decoherence on the negativity of a state.

%We generate nonclassical states of the electromagnetic field by utilizing the strong coupling between a two-level atom and a one-dimensional waveguide. 
%We study the steady-state emission from a continuously driven two-level atom using the method of quantum trajectories. We simulate the conditional evolution of a two-level atom being subjected to quadrature/homodyne measurement. From the measurement statistics we reconstruct the quantum state of the emitted field. We show that using this setup it is possible to generate coherent superpositions of Fock states beyond vacuum and single-photons. 

\end{abstract}

\maketitle

%The title needs to match abstract and introduction (and the rest)\\
%cite the paper by Pascale

%\input{lists.tex}

\section{Introduction}

%%%5%%%%%%%%%%%%%%%%%%%%%%%%%%%%%%%%%%%%%%%%%%%%%%%%%%%%%%%

Besides discrete-variable qubit-based setups, continuous-variable systems have emerged as a promising alternative for quantum information processing applications such as quantum cryptography, teleportation, and quantum computing~\cite{Lloyd1999Feb,Andersen2009Jul}. %https://journals.aps.org/prl/pdf/10.1103/PhysRevLett.102.120501
While the first two applications can be performed with Gaussian quantum states, quantum computing requires non-Gaussian states---or more specifically, quantum states with a negative Wigner function---in order to gain an advantage over classical computing~\citep{Mari2012Dec,Veitch2013Jan}. Well-known Wigner-negative states include Fock states and Schr\"odinger's cat-states. These can be created in cavities~\cite{Hofheinz2008Jul,Hofheinz2009May,Leghtas2015Feb,Touzard2018Apr} and propagating modes by controlled release~\cite{Pfaff2017Jun,Yoshikawa2013Dec,Goto2019Feb}. Special effort has been put into creation of the single-photon Fock state~\cite{Lounis2005Apr,Eisaman2011Jul,Chunnilall2014Jul} due to its usefulness for a multitude of quantum information applications, including quantum computing~\cite{Knill2001Jan,Yoran2003Jul}. Single-photon sources have been engineered for a variety of different platforms~\cite{Senellart2017Nov,He2018Jun,Darquie2005Jul,Aharonovich2011Jun,Forn-Dnaz2017Nov}. Moreover, propagating pure superpositions of vacuum, single and two-photons have been generated in superconducting circuits~\cite{Eichler2012Sep} and with quantum dots~\cite{Loredo2018Oct}. All of these setups have in common that they use pulsed excitation. We are instead interested in steady-state generation of Wigner-negative states that result from a continuous drive.

Although a method that generates steady-state Wigner-negative states of light with the help of feedback has been proposed~\cite{Joana2016Dec}, we wish to look at a much simpler system that is already experimentally available: a coherently driven two-level atom~\cite{Hoffges1998Jan,Astafiev840,Muller2007Nov}. An excited two-level system is the simplest model of a single-photon emitter. However, single-photon states cannot be generated from it in the continuous driving regime. Despite this, the resonance fluorescence emitted by this simple system is well known to exhibit nonclassical properties such as antibunching and squeezing~\cite{Kimble1977Sep,Loudon1984Feb,Schulte2015Aug}. Nevertheless, a characterization of the radiation field in terms of the Wigner function had not been performed until recently. In a previous paper, we demonstrated numerically that for certain parameter regimes, the emission from the two-level system in front of a mirror is characterized by a negative Wigner function~\cite{Quijandrna2018Dec}. Thus, this is a potential implementation for continually generating possible resource states for quantum computing. In this article, we study two possible configurations of one-dimensional resonance fluorescence: a two-level atom in a waveguide and a two-level atom in front of a mirror. Here we elaborate on the numerical methods that allow us to reconstruct the state of the emitted field. In addition, we study the effects of additional decoherence channels on the two-level system that are ubiquitous in experiments. In particular, we discuss a possible circuit-QED implementation.

The article is structured as follows. In Section~\ref{sec:wigner} we briefly introduce the Wigner function and the measure of negativity. After this, we explain the setup in Section~\ref{sec:setup}. In Section~\ref{method}, we describe our numerical methods: quantum trajectories and maximum-likelihood estimation for state reconstruction. In part~\ref{results} we show the main result: conditions under which Wigner negativity is observed. We then analyze the result in terms of coherent reflection. We also investigate how decoherence affects the negativity, and analyze the effect the purity of the state. Lastly, in Section~\ref{conclusion}, we summarize and conclude.

%

%

% photon subtracted states~\cite{Agarwal1968Jul,Neergaard-Nielsen2006Aug}

%Methods to engineer fields confined in resonators has been proposed theoretically as well as accomplished experimentally~\citep{Hofheinz2009May,Vogel1993Sep}, 

\subsection{Quantum phase space and the Wigner function}\label{sec:wigner}
%The integral over momentum gives the probability density in position, and vice versa.
The phase space formulation of quantum mechanics offers a framework where the equations of quantum mechanical systems can take the same form as classical equations of motion.
As such, the phase space formulation provides insights into the connection between classical and quantum mechanics~\cite{Hillery1984Apr,Berry1977Oct,Heller1976Aug}. There is a correspondence between quantum operators and classical functions in phase space~\cite{Polkovnikov2010Aug}; a c-number function in phase space is related to an operator in Hilbert space by the so-called Weyl correspondence, and the function is called the Weyl symbol of the operator~\cite{Blaszak2012Feb}.

In the phase space formulation, a quantum state is represented by a quasiprobability distribution. It is referred to as a \emph{quasi}probability distribution because according to the Heisenberg uncertainty principle, it is not possible to define a joint probability distribution at a point $(x,p)$ in phase space, since the corresponding operators $\hat x,\,\hat p$ do not commute~\cite{Lutkenhaus1995Apr,Hillery1984Apr}. Unlike true probability distributions, which are always positive, quasiprobability distributions can be negative in parts of phase space. This is indicative of nonclassicality~\cite{Kenfack2004Aug,Lutkenhaus1995Apr}. The Wigner function, which is the Weyl symbol of the density matrix, is such a quasiprobability distribution. We will use it in this work because its negativity has been shown to be a resource for quantum computation~\cite{Pashayan2015Aug,Rahimi-Keshari2016Jun,Albarelli2018Nov}.
%It has several nice properties; for example, it is the only quasiprobability distribution that when integrated over momentum gives the probability density in position, and vice versa.   
%The main advantage is that c-number equations are used instead of operators~\cite{Lee1995Aug}.  but also 
%nonclassical effects are associated with negative values of some quasiprobability(lutkenhaus)
%One of the benefits of this formulation of quantum mechanics is that the phase space distribution can be plotted to provide a visualization of otherwise abstract quantum states.
%Other uses of the Wigner function can be plotted to provide a visualization of otherwise abstract quantum states
In order to quantify the resourcefulness of a particular state, Ref.~\cite{Albarelli2018Nov} defines a resource monotone $\neg$ called the \emph{Wigner logarithmic negativity} (WLN)
\begin{equation}
    \mathsf{W}=\log\left(\int\lvert W(x,p)\rvert \dif x \dif p\right).
\end{equation}
It has the property $\mathsf{W}>0$ when the Wigner function $W(x,p)$ has a negative part. In our previous article~\cite{Quijandrna2018Dec}, we used the integrated negativity $\mathcal{N}=\int\left[\lvert W(x,p)\rvert-W(x,p)\right] \dif x \dif p$, which is related to the WLN by $\mathsf{W}=\log(\mathcal{N}+1)$. While it is also a monotone, the WLN has the advantage of being additive~\cite{Albarelli2018Nov}. Although additivity is not directly relevant for the purpose of this paper (any metric of Wigner negativity would suffice), we will use the WLN for the possibility to connect it to the resource theory.

%Additionally, the Wigner function is useful for quantum state tomography, which is the estimation or reconstruction of an unknown quantum state from multiple measurements on copies of said state. We will utilize this in Section~\ref{sec:trajs}. 

\subsection{Setup}\label{sec:setup}

In this setup, a two-level atom is driven by a coherent field. A coherent state is Gaussian and is thus characterized by a positive Wigner function~\cite{Hudson1974Oct}. To get a Wigner-negative state from a coherent input, a nonlinear element is required~\cite{Lloyd1999Feb}. Here the nonlinearity is provided by the two-level atom. In order to utilize the nonlinearity of the atom to create nonclassical states of light, strong coupling between the electromagnetic field and the atom is needed. One-dimensional waveguides facilitate strong coupling by confining the radiation energy in a small volume, and also avoids spatial mode mismatch between incident and scattered fields~\cite{Astafiev840}.

%https://journals.aps.org/pra/pdf/10.1103/PhysRevA.88.013855
% motivations to study TLS coupled to waveguides. Atoms having a large dipole moment (Rydberg atoms) as well as waveguides (including optical waveguides) that concentrate radiation energy in small volumes are used to increase the coupling.

We look at two different setups: a two-level atom in an infinite one-dimensional waveguide, and a semi-infinite waveguide terminated by a mirror. In our case, the distance between the atom and the mirror is considered to be negligible. This, in addition to neglecting effects from the finite size of the atom, allows for a Markovian description~\cite{Dorner2002Aug,Cao2001Sep}. 

 %the time  the  light  needs  to  bounce  back  and  forth  between  the atom and the mirror can be set essentially to zero (Markovian limit)\url{https://journals.aps.org/pra/abstract/10.1103/PhysRevA.66.023816}

%what possible configurations exist in 1D: 
%\begin{itemize}
%    \item atom with two decay channels
%    \item  an atom in front of a mirror
%    \item - an atom in front of a mirror with the mirror at a finite distance from it (this includes time delays).
%\end{itemize}

The master equation for the Markovian dynamics of an open quantum system $\hat\rho$ is (with $\hbar=1$)~\cite{breuer,wisemanbook,gardiner}:
\begin{equation}\label{ME}
    \dod{\hat\rho}{t} = -\imi[\hat H, \hat\rho]+ \sum_{k=1}^K\gamma_k\mathcal{D}[\hat L_k]\hat\rho. 
\end{equation}
%
%
%\begin{equation}\label{ME}
%    \dod{\hat\rho}{t} = -\imi[htp, \hat\rho]+ \gamma_1\mathcal{D}[\sigmaminus]\hat\rho +\gamma_2\mathcal{D}[\sigmaminus]\hat\rho, 
%\end{equation}
%
The first term on the rhs of Eq.~\eqref{ME} corresponds to unitary evolution according to the system Hamiltonian $\hat H$, and the second term describes dissipation of information into the environment, with the superoperator $\mathcal{D}$ defined as
\begin{equation}
\mathcal{D}[\hat L]\hat\rho = \hat L\hat\rho \hat L^\dagger - \frac{1}{2}\hat L^\dagger \hat L \hat\rho - \frac{1}{2}\hat\rho \hat L^\dagger \hat L.
\end{equation}
%
%nonunitary evolution, i.e. decoherence, due to interactions with the environment. These interactions are represented by a superoperator $\mathcal{D}$ and an operator $\hat L_k$ for each decay channel $k$ with rate $\gamma_k$, defined as
In Eq.~\eqref{ME} there are in total $K$ decay channels, and each channel $k$ is represented by an operator $\hat L_k$ that describes a decoherence process. We will first only look at photon emission into the waveguide, represented by the atom lowering operator $\sigmaminus$ [see Eq.~\eqref{ME2}]. Later in Section~\ref{sec:dephasing} we will also consider dephasing as well as nonradiative losses.

The atom is driven on resonance by a coherent field from one end of the waveguide. In the rotating frame of the atom, the Hamiltonian then only consists of the drive term
\begin{equation}\label{drive_hamiltonian}
    \hat H=-\imi\sqrt{\gamma_1}\Omega(\sigmaplus + \sigmaminus),
\end{equation}
%
%where $\sigmaplus$ and $\sigmaminus$ are the raising and lowering operators of the atom, 
where $\Omega$ is the drive strength, and $\sqrt{\gamma_1}$ represents the coupling to the waveguide via the channel $k=1$. The atom will emit radiation as a response to this drive. 

%where the dissipator $\mathcal{D}$ represents the amplitude decay \textcolor{red}{(use a different term? energy dissipation? didn't want to write "dissipation" twice)} channels of the system. Generally, in a one-dimensional system, there are two possible decay channels: the atom can emit a photon into the waveguide either to the left, or to the right. The decay rates for each side are $\gamma_1$ and $\gamma_2$, respectively.  \textcolor{red}{bad flow to next paragraph?}

%
%%%!!!!!!!!!!!!!!!!!!!!!!!!!!!!!!!!!!!!!!!!!!!!!!!!!!!!!
\subsubsection{Atom in an infinite one-dimensional waveguide}
\begin{figure}[hbtp!]
    \centering
    \includegraphics[width=\columnwidth]{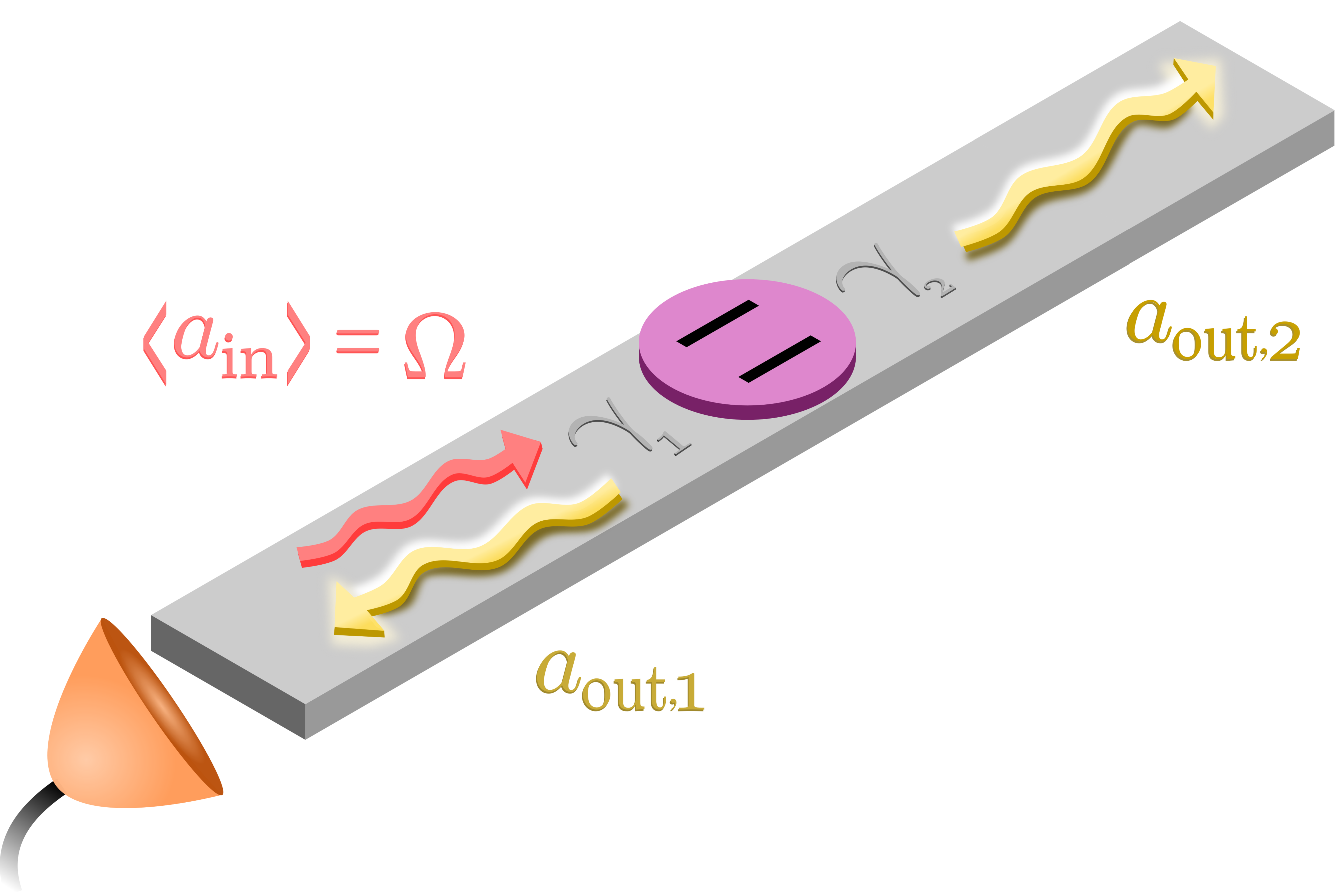}
    \caption{The two-level atom in the infinite waveguide is driven from the left side of the waveguide with strength $\Omega$. Photon emission into the left occurs with rate $\gamma_1$, and into the right with rate $\gamma_2$. Only the left-propagating radiation is detected.}
    \label{fig:open}
\end{figure}

In our one-dimensional setup, the infinite waveguide enables decay into left- and right-propagating modes, corresponding to two independent decay channels. We assign the left-going modes to channel $k=1$, and right-going to $k=2$. Driving from the left (with a right-going mode), we can consider the left-going photons as reflected and the right-going photons as transmitted, see Fig.~\ref{fig:open}. 

The master equation for a two-level atom in this setup is
\begin{equation}\label{ME2}
    \dod{\hat\rho}{t} = -\imi[\hat H, \hat\rho]+ \gamma_1\mathcal{D}[\sigmaminus]\hat\rho + \gamma_2\mathcal{D}[\sigmaminus]\hat\rho. 
\end{equation}
We look at the case when the decay rates into left-going and right-going modes are equal, that is, $\gamma_1=\gamma_2=0.5$. However, note that we only observe the left-going (reflected) field. 
%This setup has two decay channels---it can emit a photon into either side of the waveguide. 
%
%

%%%!!!!!!!!!!!!!!!!!!!!!!!!!!!!!!!!!!!!!!!!!!!!!!!!!!!!!
\subsubsection{Atom in a semi-infinite one-dimensional waveguide}

%In our case, there will be one amplitude decay channel for the semi-infinite one-dimensional waveguide; the atom can only emit photons into one end of the waveguide.

\begin{figure}[hbtp!]
    \centering
    \includegraphics[width=0.9\columnwidth]{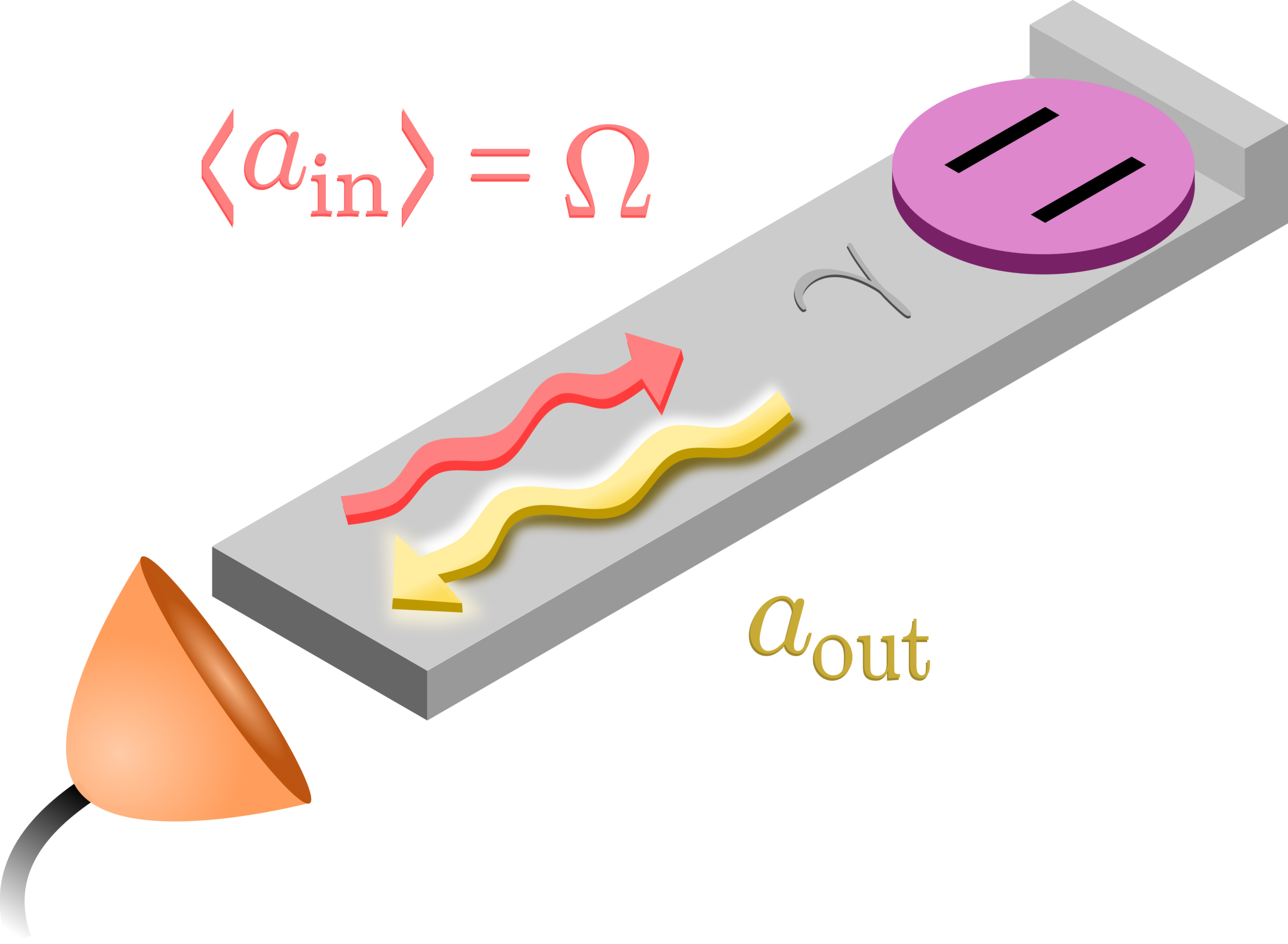}
    \caption{The two-level atom in the semi-infinite waveguide is driven with strength $\Omega$, and has a radiative decay rate $\gamma$. This setup can also be seen as an atom in front of a mirror.}
    \label{fig:mirror}
\end{figure}

We also consider a semi-infinite waveguide, ending with a reflecting boundary condition at the position of the atom. This is effectively an atom in front of a mirror, and consequently, there is only one decay channel---the same as the driving channel. See Fig.~\ref{fig:mirror}. The master equation is identical to Eq.~\eqref{ME2} but with decay rates $\gamma_2=0$ and $\gamma_1=1$. 

%Here, the previously open waveguide is terminated at one end by a mirror. This means there is only one decay channel. 
%

\subsection{Input-output formalism}

The master equation~\eqref{ME} describes the state of the two-level atom, but we are interested in the Wigner function of the state of the radiation field. The input-output relation connects the two systems~\cite{Gardinerpaper,gardiner}. In the infinite waveguide we have two output field modes: the left-going $\hat a_{\text{out},1}$ and the right-going $\hat a_{\text{out},2}$, as shown in Fig.~\ref{fig:open}. When driving from the left, the input-output relations are 
\begin{equation}\label{inout1}
  \begin{split}
  &\hat a_{\text{out},1}(t)=\sqrt{\gamma_1}\sigmaminus(t),\\
  &\hat a_{\text{out},2}(t) =\ain (t) + \sqrt{\gamma_2}\sigmaminus(t).
  \end{split}
\end{equation}
This gives a relation between the atomic lowering operator $\sigmaminus$ and the field operators $\hat a_{\text{out},i}$. 

By placing a mirror at the position of the atom, the left- and right-propagating modes are no longer independent, but coupled to each other by the input-output relation
\begin{equation}\label{inout}
    \hat a_{\text{out}}(t) = \ain (t)+ \sqrt{\gamma}\sigmaminus(t),
\end{equation}
where we have removed the subscript because there is only one decay channel. The input field corresponds to the coherent drive, with $\braket{\ain}=\Omega$.

%Interaction between the atom and the field entangles them, and observation of one will give information about the other. We will use this relation to extract information about the radiation field by solving the time evolution of the atom.

%and observing the output field has a back-action on the atom. The master equation~\eqref{ME} provides the average atom evolution,

%if the state is monitored, it will be conditioned on the measurement result due to back-action. Because quantum mechanical measurements are inherently non-deterministic, a SME is used to model the time evolution of a quantum system subject to measurements.

%but measurement of the output field have a nondeterministic effect on the atom state for every measurement. This will be further explained in the next section. \textcolor{red}{probably needs adjustment}

%\subsubsection{Resonance fluorescence}
%\input{introduction/resonance_fluorescence.tex}

%\subsection{Wigner function}
%\input{introduction/wigner.tex}

%\subsubsection{Coherent reflectance}

%\subsection{Hamiltonian etc}

%%%%%%%%%%%%%%%%%%%%%%%%%%%%%%%%%%%%%%%%%%%%%%%%%%%%%%%%%%%%%%%%%%%%%%%%%%%%%%%%%%%%%%%%%%%%%%%%%%%%%%%%%%%%%%%%%%%%%%%%%%%%%%%%%%%%%%%%%%%%%%%%%%%%%

\section{Method}\label{method}
We wish to construct the Wigner function of the reflected resonance fluorescence in steady-state ($\dot\rho=0$) in the above described setups. As stated in the previous section, we have indirect information about the state of the reflected field via the input-output equations. From this, the state of the field can be inferred by calculating an infinite hierarchy of time-ordered field correlation functions. This is clearly impractical; instead we follow a different approach, and implement one of the commonly used methods for experimental tomography. There are several ways to perform experimental Wigner tomography; the Wigner function can be directly calculated from measurements of parity~\cite{Royer1977Feb,Banaszek1999Jul,Haroche2007Sep}, field correlations~\cite{DaSilva2010Oct,Menzel2010Aug,Eichler2011Jun}, or field quadratures~\cite{Vogel1989Sep,Smithey1993Mar,DAriano2005Jul}. In the optical regime, the last example corresponds to homodyne tomography.

In this section, we first introduce quantum trajectories which allow us to simulate quadrature measurements (homodyne detection) of the resonance fluorescence to obtain artificial measurement data. We then explain how this data is used to reconstruct the density matrix of the radiation field. Subsequently, the Wigner function is calculated from the density matrix.

%Describe strategy. what do we need to calculate the wigner function of the radiation field output? why do we rely on an experimental technique? how to reproduce it numerically?

% We numerically simulate a continuously monitored two-level atom by solving a stochastic master equation. This provides so-called quantum trajectories.

%the quantum trajectory can also be interpreted as the conditional state of the system Of course the system may interact with the outside world in ways that cannot  be interpreted  as a  measurement as there is simply no  practical way of extracting information  from the environment. However, we shall consider only the more restrictive situation in which the source of irreversibility is due to a  specific measurement process.  The measurement record is a stochastic process due to the probabilistic nature of quantum measure- mentIn general, different measurement  records correspond to different  conditional states as the states will evolve  along  difficult trajectories.	~\citep{Breslin1997Nov}																				 
%Performing the inverse Radon transform is problematic because of \textcolor{red}{reasons}. Reconstruction of the density matrix using a maximum likelihood method is better \textcolor{red}{(reasons here)}.
%\input{method/homodyne.tex}

%#############################################################################################################################################
\subsection{Quantum trajectories}\label{sec:trajs}

%A quantum trajectory is the path followed by the state of a quantum system in time. In quantum trajectory simulations the instantaneous state is conditioned on the previous measurement result as a back-action. Since quantum mechanical measurements are inherently non-deterministic, the trajectory is stochastic in nature. Because of this, a stochastic master equation (SME) is used to model the time evolution of a quantum system subject to measurements. There are different SMEs corresponding to different measurement schemes, such as photon counting or homodyne detection. We look at homodyne detection because it can be reliably performed in both optical and microwave settings, as opposed to photon counting which is only viable in the optical frequencies.  

%\textcolor{red}{trajectory:realization of a stochastic process}

A quantum trajectory is the path followed by the state of a quantum system in time. If the state is continuously monitored, its time evolution will be conditioned on the measurement result due to backaction. Because quantum mechanical measurements are inherently nondeterministic, the system evolution is stochastic. For this reason, a stochastic master equation (SME) is used to model the time evolution of a quantum system subject to measurements~\cite{wisemanbook}.

%¤¤¤¤¤¤¤¤¤¤¤¤¤¤¤¤¤¤¤¤¤¤¤¤¤¤¤¤¤¤¤¤¤¤¤¤¤¤¤¤¤¤¤¤¤¤¤¤¤¤
\subsubsection{Quadrature measurement simulations}
In homodyne detection, the quadratures of an electromagnetic field represented by the bosonic creation (annihilation) operator $\adagger$ ($\ahat$) are measured continuously. For tomography purposes, the relevant observable is the generalized quadrature~\cite{Leonhardt1997Jul}
%We numerically simulate a two-level atom subject to continuous weak measurements of the output field quadratures. The relevant observable is the generalized quadrature~\cite{Leonhardt1997Jul}
%
\begin{equation}\label{quads}
     \xhattheta=\frac{1}{\sqrt 2}(\adagger\expup{\imi\theta}+\ahat\,\expup{-\imi\theta}) =  \hat X\cos \theta +  \hat P \sin \theta,
\end{equation}
where $\hat X$ and $\hat P$ are the canonically conjugate position and momentum operators 
\begin{equation}\label{bothquads}
\begin{split}
     \hat X=& \frac{1}{\sqrt{2}}\left(\adagger + \ahat\right), \\
     \hat P=& \frac{\imi}{\sqrt{2}}\left(\adagger - \ahat\right),
\end{split}
\end{equation}
%%, and it is set by the phase of a local oscillator~\citep{Breitenbach1997Nov}. 
whose corresponding Weyl symbols $x,\,p$ span phase space. The parameter $\theta$ selects which quadrature is measured, and in experimental realizations its value is set by the phase of a local oscillator~\citep{Breitenbach1997Nov}. %The SME corresponding to quadrature measurements is~\cite{wisemanbook} 

The most general setup considered here corresponds to a two-level atom which decoheres through $K$ channels represented by operators $\hat L_k$ [see the master equation~\eqref{ME}]. However, only one channel is monitored in our setup. The conditional state $\hat\rho_c$ which results when only the first channel ($k=1$) is observed is given by~\cite{wisemanbook}
%
%\begin{equation}\label{sme}
%     \dif\rhoc = -\frac{\imi}{\hbar}[\hd, \rhoc]\dt + \gamma\mathcal{D}[\sigmaminus]\rhoc\dt + \sqrt{\gamma_1}\mathcal{H}[\expup{-\imi\theta}\sigmaminus]\rhoc\dif W,
%\end{equation}
%
\begin{equation}\label{sme}
\begin{split}
     \dif\rhoc = &-\imi [\hat H, \rhoc]\dt + \sum_{k=1}^K\gamma_k\mathcal{D}[ \hat L_k]\rhoc\dt + \\
     &+ \sqrt{\gamma_1}\mathcal{H}[\expup{-\imi\theta} \hat L_1]\rhoc\dif W,
     \end{split}
\end{equation}
%A quantum trajectory is the path followed by the state of a quantum system in time. If the state is monitored, it will be conditioned on the measurement result due to back-action. Because quantum mechanical measurements are inherently non-deterministic, a SME is used to model the time evolution of a quantum system subject to measurements. There are different SMEs corresponding to different measurement schemes, such as photon counting or homodyne detection~\cite{wisemanbook}. % We use homodyne detection because it can be reliably performed in both optical and microwave settings, as opposed to photon counting which is only viable for optical frequencies. 

%During homodyne measurements the signal indented to be measured is mixed with the signal of a local oscillator with phase $\theta$. 
where the measurement superoperator $\mathcal{H}$ is~\cite{Jacobs2006Sep}
\begin{equation}\label{super_meas}
     \mathcal{H}[\hat L]\hat\rho = \hat L\hat\rho+\hat\rho \hat L^\dagger - \braket{\hat L+\hat L^\dagger}\hat\rho.
\end{equation}
The stochastic nature of the measurement is provided by the Gaussian random variable $\dif W$ which has variance $\dif t$ and ensemble average $\mathds{E}[\dif W]=0$. Note that the latter property ensures that the ensemble average over trajectories correspond to the nonconditional master equation~\eqref{ME}, and the unconditional state $\rho=\mathds{E}[\rhoc]$ is the average of an ensemble of conditional states for different trajectories. One solution of the SME~\eqref{sme} corresponds to one quantum trajectory. 

The simulated measurement signal associated with~\eqref{sme} for $\hat L_1=\sigmaminus$, corresponding to the photocurrent in optical homodyne detection, is
\begin{equation}\label{signal}
    \dif j(t) \dif t= \frac{1}{\sqrt{2}}\left(\sqrt{\gamma_1}\braket{\sigmaplus\expup{\imi\theta} + \sigmaminus\expup{-\imi\theta}}\dt + \dif W\right).
\end{equation}
The deterministic part of~\eqref{signal} is directly proportional to the generalized quadrature of the continuous field $\aout$ through the input-output relation. We now explain how a bosonic mode $\ahat$ can be selected out of the continuum of modes described by $\aout (t)$.
% between the output field $\aout$ and the atomic operator $\sigmaminus$, and the relation~\eqref{propermode} between $\aout$ and the mode $\hat A$.

%A single trajectory, corresponding to a measurement record of the homodyne current, can be simulated by solving the SME~\eqref{sme} 

%&&&&&&&&&&&&&&&&&&&&&&&&&&&&&&&&&&&&&&&&&&&&&&&&&&&&&&&&&&&&&&&&&&&&&&&
%Choosing a temporal mode indeed realizes a single-mode measurement: the state is thus traced over all the othermodes, and the process would result in a statistical mixtureif there were some correlations with other modes https://journals.aps.org/prl/pdf/10.1103/PhysRevLett.111.213602

\subsubsection{Mode selection}

%\textcolor{red}{The  single mode can  be  isolated  from  the  continuum  of  modes  by performing  temporal  mode  matching,  i.e.,  integrating  the continuous signal over the temporal profile of the photon pulse which is to be characterized https://journals.aps.org/pra/pdf/10.1103/PhysRevA.86.032106}

Because the output field is not confined to a cavity, there are no discrete eigenmodes, instead we have continuous-mode field operators $\hat a_{\text{out},k}(t)$ that obey the commutation relation $[\hat a_{\text{out},k}(t),a^\dagger_{\text{out},k'}(t')]=\delta_{kk'}\delta(t-t')$~\cite{Blow1990Oct}.
%\textcolor{red}{monochromatic creation operator, single frequency mode}
%
%\begin{equation}\label{cont_time_mode}
%    a(t)=\frac{1}{\sqrt{2\pi}}\int_{-\infty}^\infty \dif \omega a(\omega) \expup{-\imi\omega t}.
%\end{equation}
%
%The photon flux is given by $n(t)=\adagger(t)a(t)$. 
To get a well-defined state containing a finite number of photons, a continuous mode must be filtered to create a wavepacket~\cite{loudon}. A mode function $f(t)$ defines the temporal profile of the wavepacket~\cite{Brecht2015Oct}. 
%A photon temporal mode is chosen by means of a temporal mode function $f(t)$. It defines the temporal shape of the wavepacket~\cite{Brecht2015Oct}.
The creation operator for a photon wavepacket in a particular temporal mode $f$ is 
% wavepacket mode function f(t)
%
\begin{equation}\label{propermode}
    \hat A^\dagger_f=\int_{0}^\infty f(t) \aout^\dagger(t) \dif t.
\end{equation}
The mode function must be normalized, i.e. $\int_0^\infty |f(t)|^2 \dt = 1$, for the bosonic mode $ \hat A_f$ to fulfill the commutation relation $[\hat A_f,\hat A_f^\dagger]=1$. 

%The probability to detect a photon at time $t$ is given by $\lvert f(t)\rvert^2$. 
Our choice of mode function is based on the fact that we monitor the steady-state output, and do not want to introduce any time-dependence. For this reason, we use a simple boxcar filter %~\cite{Woolley2013Oct} for prob
\begin{equation}\label{eq:boxcar}
f(t) = \frac{1}{\sqrt{T}} \left[ \Theta(t-t_0) - \Theta(t-t_0 -T) \right],
\end{equation}
where $\Theta(t)$ is the Heaviside step function, $T$ is the duration of the measurement, and $t_0$ is the time when the measurement starts. The filter is constant $1/\sqrt{T}$ within the time interval $[t_0,t_0+T]$ and zero outside it. 
%
%\begin{equation}
%    A^\dagger_f=\frac{1}{\sqrt{T}}\int_0^T  \adagger(t) \dif t
%\end{equation}
%
%\textcolor{red}{will I even use this???}
%&&&&&&&&&&&&&&&&&&&&&&&&&&&&&&&&&&&&&&&&&&&&&&&&&&&&&&&&&&&&&&&&&&&&&&
Because the boxcar filter~\eqref{eq:boxcar} is a real function, we can directly filter the photocurrent $j(t)$ to obtain the quadratures~\eqref{quads} of $A_f$. 
In the numerical implementation, the filtering and integration simply amounts to to a summation over a subset of the time steps in the simulation:
\begin{equation}\label{eq:photocurrent}
    J = \int \dif j(t)f(t) \rightarrow \sum_{i=n_0}^n \frac{j_i}{\sqrt T}.
\end{equation}
We first let the system evolve without recording the signal $j_i$ from time $t=0$ to $t=t_0$ where it has reached steady state. With a time discretization $\dif t$, this corresponds to time step $n_0=t_0/\dif t$. After this, the system evolves until time $t=t_0+T$ while the signal is integrated, finishing at step $n=(t_0+T)/\dif t$. 

There is one integrated signal $J$ per trajectory. To reconstruct a quantum state, repeated measurements on a large ensemble of identically prepared states must be performed. This amounts to simulating multiple trajectories. The many integrated photocurrents, which represent the quadrature values plus noise, are recorded and sorted into equally sized bins to create a measurement histogram for each value of $\theta$. The histograms are then used for the maximum likelihood reconstruction of the density matrix.

%The binning amounts to discretization of the continuous variable quadrature values.  

%Thus, the total number of time steps in the simulation is $n=(t_0+T)/\dif t$, and the photocurrent summation commences at step $n_0=t_0/\dif t$. The integrated signal is recorded for each simulated trajectory. 

%

%#############################################################################################################################################
\subsection{Maximum likelihood state reconstruction}\label{sec:maxlik}
%Quantum state tomography, or quantum state estimation

%\textcolor{red}{easy way to get error statistics, take $|\rho-\rho|$. \citep{Lvovsky2009Mar} \url{http://iopscience.iop.org/article/10.1088/1367-2630/10/4/043022/meta} apparently says it's worse than the Fisher matrix. read that}

%\begin{itemize}
%    \item To reconstruct a quantum state, repeated measurements on a large ensemble of identically prepared states must be performed. 
        %\item Histogram = marginal distribution Pr($x_\theta)$
%\end{itemize}

%other methods needs prior knowledge about the system \citep{Mogilevtsev1997Nov} maxlik: need no prior information about the reconstructed state

%We show in the appendix that the probability density of a given quadrature is the integral of the W function along a line in phase space orthogonal to the direction of this quadrature. Measuring the quadrature fluctuations for all possible phases thus amounts to determining the integrals of W along all possible directions in the phase plane. haroche exploring the quantum

For a system in state $\hat\rho$, the measurement histogram of the observable $\xhattheta$ approximates the probability 
\begin{equation}\label{eq:pr}
    \text{Pr}(\xtheta)=\braket{\xtheta|\hat\rho|\xtheta} = \tr[\hat\Pi^\theta \hat\rho], 
\end{equation}
of detecting the associated eigenvalue $\xtheta$, defined by $\xhattheta\ket{\xtheta}=\xtheta\ket{\xtheta}$~\cite{wisemanbook}. In the limit of an infinite number of measurements (trajectories) the histogram is identical to~\eqref{eq:pr}. We have also defined $\hat\Pi^\theta =\ketbra{x_\theta}{x_\theta}$, which is the projector onto the quadrature eigenstate $\ket{\xtheta}$. 

Our aim is to reconstruct the density matrix $\hat\rho$ of the radiation field in the Fock basis $\{\,\ket{n}\}$.  In this basis, the projector $\hat\Pi^\theta$ has matrix elements
\begin{equation}\label{projs_mn}
    \Pi^\theta_{mn}=\braket{m|\xtheta}\braket{\xtheta|n}=\psi^*_m(\xtheta)\psi_m(\xtheta),
\end{equation}
where $\psi_n(\xtheta)$ is the $n$th harmonic oscillator eigenfunction in the position basis, multiplied by an additional phase factor $\exp(-\imi n \theta)$~\cite{Lvovsky2004May}. For each $\theta$, the quadrature $\xhattheta$ is a continuous variable operator with eigenvalues on the real axis ($\xtheta\in\mathbb{R}$). In order to construct a measurement histogram it is necessary to discretize a region of the real axis into a finite number of bins. The probability of observing $\xtheta$ in bin $j$ is given by 
\begin{equation}\label{pr_k}
    \text{Pr}(\xtheta,j)=\tr[\hat\Pi^{\theta,j}\hat\rho]
\end{equation}
where the projector $\hat \Pi^{\theta,j}$ has Fock basis matrix elements obtained by integrating~\eqref{projs_mn} over histogram bin $j$:
\begin{equation}\label{eq:pi_int}
   \Pi_{mn}^{\theta,j}  = \int_{x_{\theta,j}}^{x_{\theta,j+1}} \psi_m(x_\theta)^* \psi_n(x_\theta) \dif x_\theta.
\end{equation}

\begin{figure}[hbt!]
    \centering
    \includegraphics[width=\columnwidth]{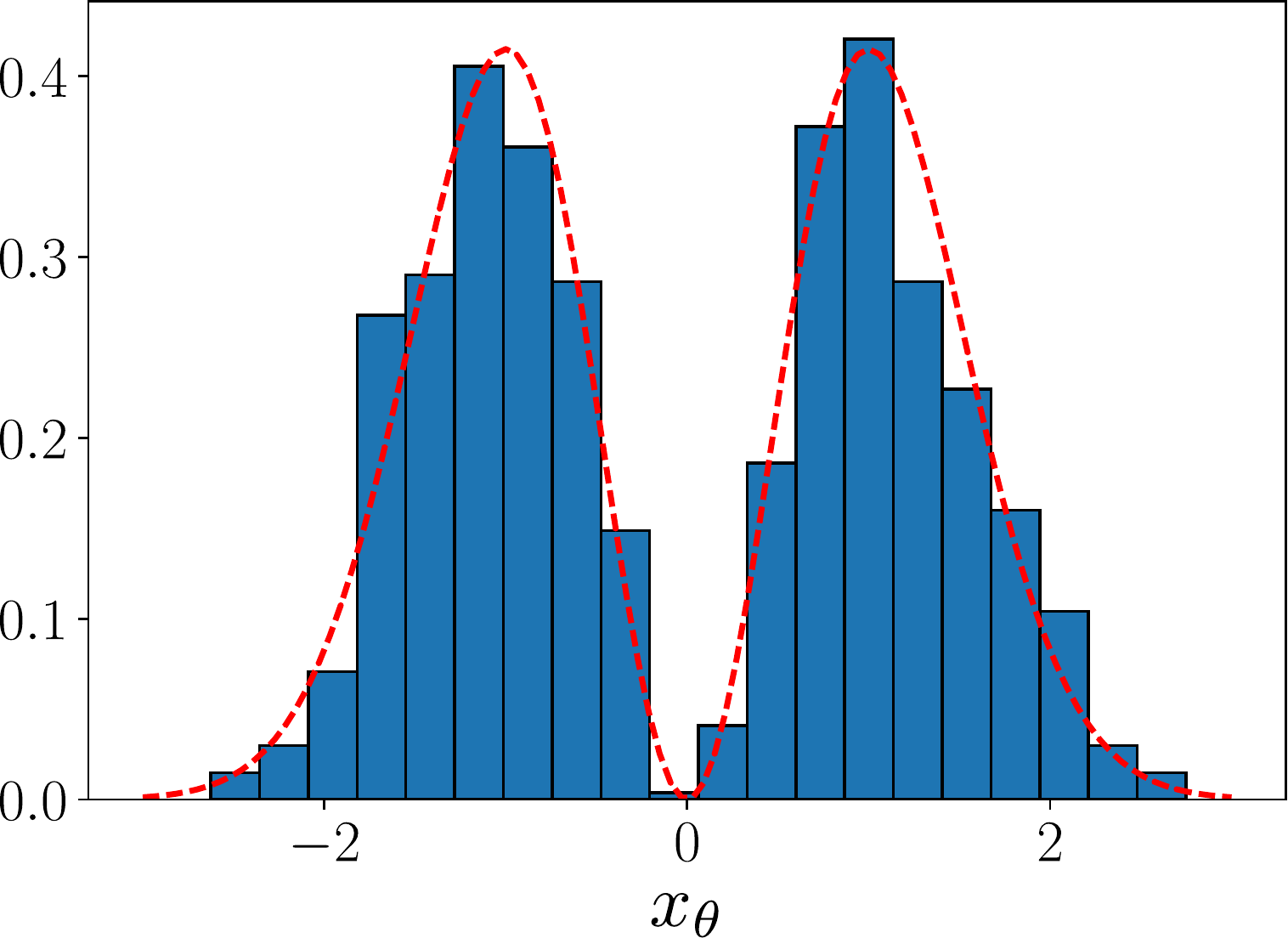}   
    \caption{Normalized histogram of the integrated signals from 1000 quantum trajectories of an initially excited two-level atom by using Eq.~\eqref{sme} with zero drive. The dashed line corresponds to the theoretical probability $\text{Pr}(\xtheta)=\lvert \psi_1(x_\theta)\rvert^2$ for a field in a single-photon state. }
    \label{fig:hist}
\end{figure}

A measurement histogram contains $n_{\theta,j}$ counts per bin $j$ for a particular phase $\theta$. The corresponding normalized histogram is given by $n_{\theta,j}/N$, with $N=\sum_j n_{\theta,j}$ the total number of counts. As an example, in Fig.~\ref{fig:hist}, we plot the normalized histogram for a single-photon state $\hat\rho=\ketbra{1}{1}$. This state is spherically symmetric in phase space, which means that the histogram is independent of $\theta$. For a pure single-photon state, the probability to observe quadrature $\xtheta$ is $\text{Pr}(\xtheta)=\lvert \psi_1(x_\theta)\rvert^2$; the squared amplitude of the first excited harmonic oscillator wavefunction. Fig.~\ref{fig:hist} shows that the normalized measurement histogram indeed approaches $\lvert \psi_1(\xtheta)\rvert^2$. 
%an observable $\mathcal{O}$ corresponds to the probability of observing its different eigenvalues $\lambda$. The observable has a spectral decomposition $\mathcal{O}=\sum_\lambda \lambda \Pi_\lambda$
%
%
%where we have defined $\Pi_\theta =\ketbra{x_\theta}{x_\theta}$, a projector onto quadrature eigenstates $\ket{\xtheta}$, associated with a measurement of quadrature $\xtheta$.

%%%%%

Since the state $\hat\rho$ determines the measurement statistics of $\hat\Pi^\theta$, information about the underlying quantum state can be extracted from quadrature measurement histograms. A histogram is an approximation of $\text{Pr}(\xtheta)$, which is in fact a projection of the integrated Wigner function on a plane in phase space orthogonal to the measured quadrature. In other words, the integral of the Wigner function over a certain quadrature $x_{\theta+\pi/2}$ gives the probability distribution of measuring the conjugate quadrature $x_\theta$~\cite{Serafini2017Jul}. The Wigner function is the only quasiprobability distribution with this property~\cite{Haroche:993568}. 
%
%\begin{figure}[h!]
%    \centering
%    \includegraphics[width=\columnwidth]{bilder/hist-crop.pdf}
%    \caption{The measurement histograms are projections of the integrated Wigner function $W$ on a plane in phase space. To reconstruct the Wigner function from measurement histograms, measurements are taken for a number of different phase angles $\theta$.}
%    \label{fig:hist_proj}
%\end{figure}

However, directly calculating the Wigner function from the histograms is fraught with numerical difficulties; even small errors can lead to inaccurate and even unphysical features of the corresponding density matrix: its diagonal elements may be found to be negative, and it is not guaranteed to have trace one~\cite{Lvovsky2009Mar}. A more robust method is to use \emph{maximum-likelihood estimation}~\cite{Hradil1997Mar} to reconstruct the density matrix, from which the Wigner function then can be calculated. %It is an iterative method for finding the density matrix $\hat\rho$ that maximizes the likelihood of having observed a particular set of measurement data.   
Maximum-likelihood estimation of a quantum state is a method of statistical inference for finding the density matrix $\hat\rho$ that maximizes the likelihood that, given a particular set of measurement histograms represented by $n_{\theta,j}$, the system was prepared in state $\hat\rho$~\cite{Hradil2019}. The likelihood functional to be maximized is defined as
\begin{equation}\label{likelihood}
    \mathcal{L}(\hat\rho)=\prod_{j,\theta}\tr[\hat\Pi^{\theta,j}\hat\rho]^{ n_{\theta,j}}.
    \end{equation}
Maximizing the likelihood~\eqref{likelihood} is equivalent to minimizing the statistical distance between (normalized) measurement data $n_{\theta,j}$ and probabilities~\eqref{pr_k} predicted from the quantum state $\hat\rho$~\cite{Hradil2019}.

%For each $\theta$, the quadrature $\xhattheta$ is a continuous variable with eigenvalues on the real axis ($\xtheta\in\mathbb{R}$). In order to construct a measurement histogram it was necessary to discretize a region of the real axis into a finite number of (equally sized) bins. 

%The probability of observing $\xtheta$ in bin $k$ ($[x_{\theta,k},x_{\theta,k+1}]$) is given by $\text{Pr}(\xtheta,k)=\tr[\hat\Pi^{\theta,k}\hat\rho]$, with the projector for bin $k$ having elements

The matrix elements of the projectors $\hat\Pi^{\theta,j}$ were obtained by evaluating the integrals~\eqref{eq:pi_int} numerically with the trapezoidal rule. These projectors, along with the histogram data $n_{\theta,j}$, was used to reconstruct the maximum-likelihood density matrix with the iterative method presented in Ref.~\cite{Lvovsky2004May}. The initial density matrix was chosen to be the normalized identity matrix, and we stopped the iterations when the Frobenius norm $\|A\|_\text{F}=\sqrt{\tr(\hat A\hat A^\dagger)}$ of the difference between two consecutive density matrices was less than $ 10^{-6}$. To get sufficient tomographic data, the phase angle $\theta$ was varied between 0 and $90\degree$~\citep{Leonhardt1997Jul}, divided into 20 increments of $4.5\degree$. We ran between 500 and 1000 trajectories per phase $\theta$, and used histograms with 100 bins over the range $\xtheta\in[-5,5]$, which was suitable for the explored parameter regimes of drive strength $\Omega$ and integration time $T$. A link to the code used for the trajectory simulations and maximum-likelihood estimations can be found in Ref.~\cite{ingrid_strandberg_2019_3371360}.

When the density matrix $\hat\rho$ of the radiation field in the Fock basis has been obtained, the corresponding Wigner function is calculated numerically as
\begin{equation}
  W(x,p) = \sum_{mn} \hat\rho_{mn} W_{mn}(x,p),
\end{equation}
where the expression for $W_{mn}$ is given in Appendix~\ref{sec:app1}.

%where $m$ and $n$ go from zero up to the selected Fock space dimension.

%%%%%%%%%%%%%%%%%%%%%%%%%%%%%%%%%%%%%%%%%%%%%%%%%%%%%%%%%%%%%%%%%%%%%%%%%%%%%%%%%%%%%%%%%%%%%%%%%%%%%%%%%%%%%%%%%%%%%%%%%%%%%%%%%%%%%%%%%%%%%%%%%%%%%
\section{Results and discussion}\label{results}
In this Section we present and explain the results of the numerical study. First we present the main result---the absence and presence of Wigner negativity in the resonance fluorescence for the infinite and semi-infinite waveguides, respectively. We find the optimal drive strength $\Omega$ and integration time $T$ to maximize the WLN, and explain this value of $\Omega$ by minimizing the coherent part of the resonance fluorescence. We then describe how the purity of the state influences the Wigner negativity. Finally, we look at the effects on the WLN that comes from adding additional decoherence channels due to pure dephasing and nonradiative decay through coupling between the atom and the environment.

%\textcolor{blue}{first contribution is from single photon (relatively weak drive}
\subsection{Wigner negativity}
\subsubsection{Atom in an infinite one-dimensional waveguide}\label{sec:infinite_waveguide}

%\textcolor{red}{arguments
%\begin{itemize}
%    \item purity vs. N
%    \item single photon is most negative
%    \item more vacuum
%    \item observations
%\end{itemize}
%}

In this setup we observe no Wigner negativity for any parameter settings. A way to understand this is that the presence of an unobserved decay channel leads to an additional admixture of vacuum into the state (the vacuum state is Gaussian and has a positive Wigner function). Imagine we place an excited two-level atom in a 1D waveguide and let it decay. The atom relaxes to the ground state by emitting a photon. If there is only one decay channel, this spontaneous emission has a well-defined temporal profile: exponential decay. Using this profile as the temporal mode function for the homodyne mode selection allows reconstruction of the single photon~\cite{Lvovsky2001Jul,Eichler2011Jun}.

However, in the infinite waveguide, there are two decay channels with decay rates $\gamma_1$ and $\gamma_2$. The probability to observe a single photon in decay channel 1 is $\rho_1=\gamma_1/(\gamma_1+\gamma_2)$. When monitoring only this end of the waveguide, regardless of the choice of temporal mode function, the single-photon state cannot be reconstructed. The reason for this is loss of information due to the nonzero probability of decay into the other end of the waveguide. In other words, the possibility of decay into channel 2 will make the observed state a statistical mixture of a single photon and vacuum, the latter with probability $\rho_0=\gamma_2/(\gamma_1+\gamma_2)$.

The Wigner function for a statistical mixture in the $\{\ket 0,\,\ket 1\}$ subspace is $W=\rho_0 W_0 + \rho_1 W_1$, where $W_0$, $W_1$ are the Wigner functions for the vacuum and single photon states, respectively. Its analytical form is~\cite{Case2008Sep}
\begin{equation}\label{wigner_10}
\begin{split}
  W(x,p)&=  \frac{1}{\pi}\expup{-(x^2+p^2)}[\rho_0 - \rho_1 + 2\rho_1(x^2+p^2) ] = \\
   & = \frac{1}{\pi}\expup{-(x^2+p^2)}[1 + 2\rho_1(x^2+p^2-1) ],
  \end{split}
\end{equation}
where $\rho_0+\rho_1=1$ was used for the second equality. We can see from Eq.~\eqref{wigner_10} that the maximum negativity occurs at the origin, and the condition for negativity to be present is $\rho_1>0.5$. In our setup, we have equal decay rates $\gamma_1=\gamma_2$, meaning there is an equal probability for the excited atom to emit a photon into either end of the waveguide. This gives gives $\rho_0=\rho_1=0.5$ and the strict inequality is unfulfilled, meaning that the Wigner function will always be nonnegative for an excited atom decaying into this infinite waveguide.

In the case of a continuous drive, in addition to the loss of information into the unmonitored decay channel, there is an additional vacuum contribution resulting from the drive itself. An initially excited two-level atom emits a single photon in a well-defined temporal profile (exponential decay). As mentioned, this allows for perfect mode matching and thus reconstruction of the single-photon state. However, when driven continuously, the atom is repeatedly excited and deexcited through both spontaneous and stimulated emission. This adds uncertainty in the time of photon emission such that there is no single well-defined temporal mode function for emission into the waveguide. Failure to exactly mode-match the single-photon state leads to additional vacuum noise~\cite{Ou1995Oct,Lvovsky2009Mar}, suggesting that the case with a continuous drive will produce an observed quantum state with a larger vacuum contribution than the case with the simply excited atom. This hints toward only Wigner-positive states being attainable from this setup.

%having an appropriate mode function $f(t)$ is essential in order to remove vacuum noise~\cite{Ogawa2016Jun}. 

%An additional complication is that our states are not necessarily restricted to the $\{\ket{0},\,\ket 1\}$ subspace, which complicates the analysis. But with the same parameters that maximizes the WLN in the semi-infinite waveguide, the probability of having more than one photon in the infinite transmission line is negligible, as can be seen in Fig~\ref{pops}. %The probability of vacuum is predominate for this setup, as opposed to the single-photon probability for the semi-infinite waveguide.
%Comparing to the case with a semi-infinite waveguide%  which is higher than that of the Wigner-negative state in Fig~\ref{wigner_1chan} which has a purity of 0.69.\textcolor{red}{and it is very pure because it is mostly vacuum?}
%Purity  for $\Omega=0.35$ $T=4$, $P=0.89$

\subsubsection{Atom in a semi-infinite one-dimensional waveguide}\label{sec:semi-infinite}

\begin{figure}[hbt!]
    \centering 
    % exjobb_program/map/plot_onecolormap_seaborn_paper.py
    \includegraphics[width=\columnwidth]{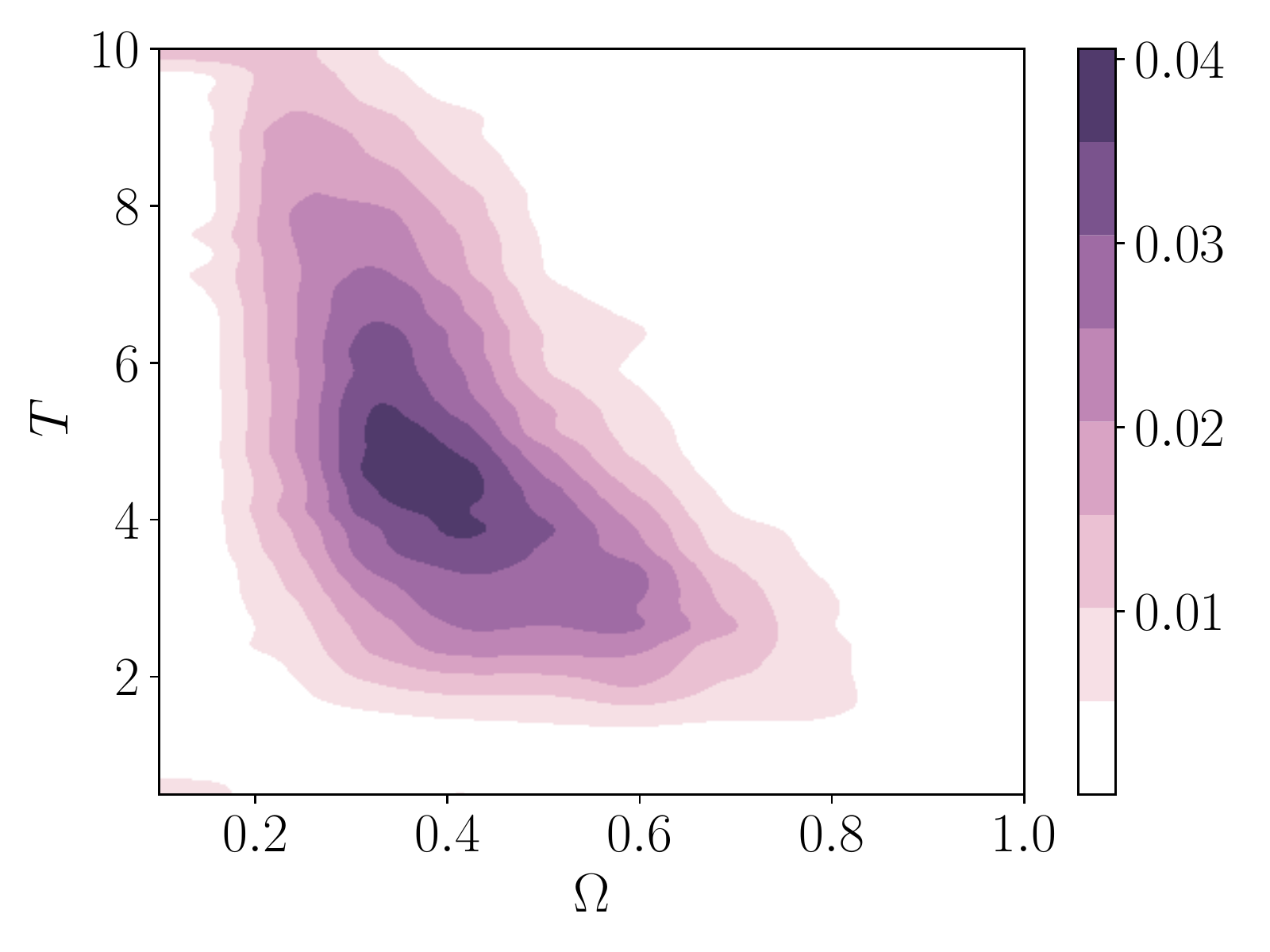}
    \caption{The WLN as a function of drive strength $\Omega$ and integration time $T$. The distance between time points is 0.2, and for $\Omega$ the spacing is 0.1. The unit is $\gamma=1$. A Gaussian interpolation is used to smoothen the data. }
    \label{fig:map}
\end{figure}

With the atom in a in a semi-infinite one-dimensional waveguide, we observe Wigner negativity for a range of parameter combinations $(\Omega,T)$. Fig.~\ref{fig:map} shows a map of the WLN as a function of $(\Omega,T)$. The decay rate is fixed to $\gamma=1$.
%p

\begin{figure}[hbt!]
% ~/exjobb_program/NEWDATA2/average_W.py
    \centering
        \includegraphics[width=\columnwidth]{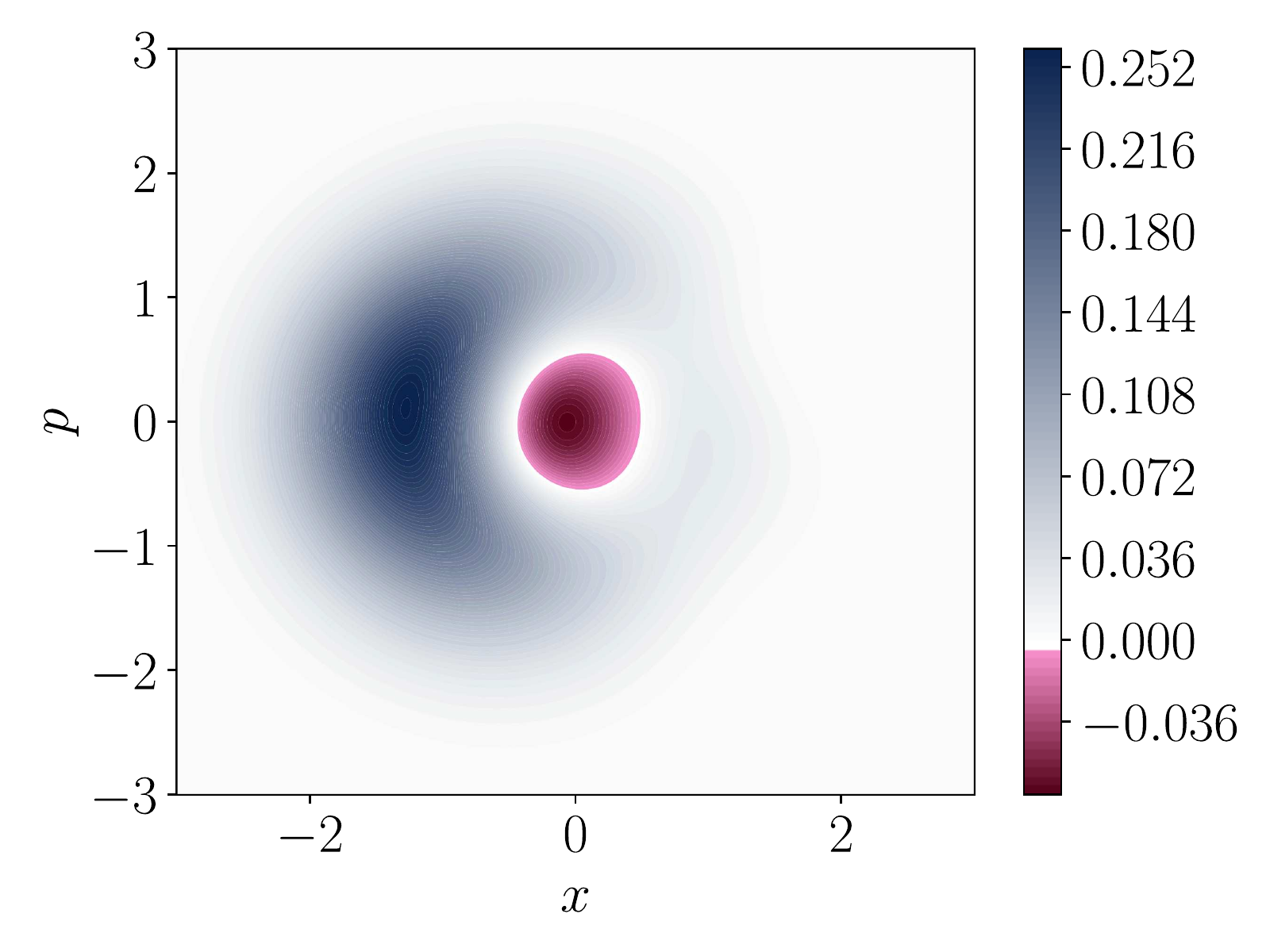}
        \caption{Wigner function of the reconstructed state for $T=4,\, \Omega=1/\sqrt{8}$ for the atom in a semi-infinite waveguide, with unit $\gamma=1$. This is the most negative state produced with this setup. The density matrix populations of this state are displayed in Fig~\ref{pops}.
        }
        \label{wigner_1chan}
\end{figure}

The maximum WLN occurs for $T$ around 4 and $\Omega=1/\sqrt{8}\simeq 0.35$. The Wigner function for the corresponding state is displayed in Fig.~\ref{wigner_1chan}, where the negative part is clearly visible. %The photon populations can be seen in Fig.~\ref{pops}, where they are compared to the populations with the infinite waveguide. We see that there is a significant single-photon population, which likely contributes to the negativity.
Why the particular integration time $T=4$ is favorable is not known to us, but the optimal drive strength $\Omega$ can be understood in terms of coherent reflectance: the highest WLN appears when the coherent reflectance is zero. This is further explained in the following section. %Section~\ref{sec:coherent_refl}.

%With the mirror, the output photon flux $\braket{\adagout\aout}$ is always equal to the input flux $\braket{\adagin\ain}=\Omega^2$. 

%¤¤¤¤¤¤¤¤¤¤¤¤¤¤¤¤¤¤¤¤¤¤¤¤¤¤¤¤¤¤¤¤¤¤¤¤¤¤¤¤¤¤¤¤¤¤¤¤¤¤¤¤¤¤¤¤¤¤¤¤¤¤¤¤¤¤¤¤¤¤¤¤¤¤¤¤¤¤¤¤¤¤¤¤¤¤¤¤¤¤¤¤¤¤4

\subsection{Coherent reflectance}\label{sec:coherent_refl}

In classical radiation theory, when an atom is irradiated, dipole oscillations are induced in the atom which in turn reemits the light at the same frequency and phase as the drive field. This is elastic and coherent scattering. In the quantum theory, light scattered of a two-level atom consists of two contributions: the first one is the coherently scattered field, given by the average dipole moment $\braket{\sigmaminus}$, which is proportional to the scattered field amplitude $\braket{\aout}$. But quantum fluctuations also have to be taken into account. The second contribution, which is incoherent in the sense that there is no fixed phase relation between the drive and scattered field, comes from spontaneous emission which occurs when the excited atom interacts with vacuum fluctuations of the surrounding electromagnetic field~\cite{cohen1998atom}. To alleviate the notation, we now drop the subscript for the output field $\aout$. The fluctuating field can be written as $\delta \hat a=\hat a-\braket{\hat a}$. The number of incoherent photons $\ninc$ is then related to the fluctuations by $n_\text{inc}=\braket{\delta \hat a^\dagger \delta \hat a}=\braket{\hat a^\dagger \hat a}-\braket{\hat a^\dagger}\braket{\hat a}$. When $\ninc=0$, the first-order correlation factorizes: $\braket{\hat a^\dagger \hat a}=\braket{\hat a^\dagger}\braket{\hat a}$. This is the definition of a first-order coherent field~\cite{Glauber2007Jan,loudon}. With this definition we characterize the incoherent part of the emission as the part of the emission which is not first-order coherent. The coherent part of the system response corresponds to the time-independent contribution from the steady-state two-time correlation function $\braket{\hat a^\dagger(t) \hat a(0)}$, as calculated in Ref.~\cite{Quijandrna2018Dec}. The same result can also be achieved without having to calculate the two-time correlation, the field expectation value $\braket{\hat a}$ will suffice. To do this, we define the coherent reflectance $r$. 
%\subsubsection{Resonance fluorescence}\label{sec:res_fluor}
%In general, when the $n$th order correlation function factorizes the field is said to be $n$th order coherent \textcolor{red}{relevant?}.
%\textcolor{red}{average dipole moment $\braket{\sigmaminus}$, steady state $\ss{\sigmaminus}$}. Fluctuations away from steady state can occur . Incoherent component arising from quantum fluctuations. Common to calculate the coherent and incoherent parts of the spectrum
%Carmichael an open systems approach

%\subsubsection{Coherent reflectance \textcolor{red}{subsection names?}}
Since we have continuous modes, the quantity we will look at is the photon flux $\hat n(t)=\hat a^\dagger(t)\hat a(t)$, rather than the photon number. But for simplicity, we omit the time argument in the following. 

The total reflected power $R$ is the ratio between the output and input flux. When the input is a coherent field with mean intensity $\Omega^2$, we have
%Spontaneous emission occurs when the vacuum field interacts when the excited atom, causing quantum fluctuations. 
%The scattered field consists of two contributions: the first one is the coherently scattered field, and is given by the mean field amplitude $\braket{a}$. The second one is the incoherently scattered field, which is spontaneous emission. Spontaneous emission occurs when the vacuum field interacts when the excited atom, causing quantum fluctuations. 
\begin{equation}
       R=\frac{\braket{\asmalldag \asmallhat}}{\braket{\adagin\ain}}= \frac{\braket{\asmalldag \asmallhat}}{\Omega^2}.
\end{equation}
We can split the output photon flux $\braket{\asmalldag \asmallhat}$ into its incoherent and coherent parts, as described above:
\begin{equation}
    R=\frac{\ninc+\ncoh}{\Omega^2}=\frac{n_\text{inc}}{\Omega^2} + \frac{\lvert\braket{\hat a}\rvert^2}{\Omega^2}=\frac{n_\text{inc}}{\Omega^2}+r^2
\end{equation}
where we define
%
%\begin{equation}
%    r=\left\lvert\frac{\braket{a}}{\Omega}\right\rvert
%\end{equation}
%
\begin{equation}
    r=\frac{\lvert\braket{\hat a}\rvert}{\Omega}
\end{equation}
as the \emph{coherent reflectance}. If $r=0$, which occurs when $\braket{\hat a}=0$, the reflected field is entirely incoherent. We call the drive strength $\Omega$ for which $r=0$ the \emph{incoherent drive point}. If $r=1$, the reflection is instead fully coherent.

To calculate the coherent reflectance, we utilize the input-output relation~\eqref{inout} to get
\begin{equation}\label{eq:r_eq}
    r=\frac{\lvert\Omega+\sqrt{\gamma}\braket{\sigmaminus}\rvert}{\Omega}.
\end{equation}
Since we are interested in the steady state, we solve $\ss{\sigmaminus}$ from the equation $\braket{\dot{\sigma}_-}=\tr[\sigmaminus\dot\rho]$ with $\dot\rho=0$, using the master equation~\eqref{ME}. The result is
\begin{equation}
    \braket{\sigmaminus}=-\frac{2\Omega/\sqrt{\gamma}}{1+8\Omega^2\gamma},
\end{equation}
and inserting this in Eq.~\eqref{eq:r_eq} gives the coherent reflectance
\begin{equation}
    r=\lvert1-\frac{2}{8\Omega^2+\gamma}\rvert.
\end{equation}
When $r=0$, the coherent part of the reflected field is zero due to destructive interference between the field reflected by the atom and the field reflected by the 
mirror~\cite{hoi}. Solving for the driving power where $r=0$ gives $\Omega=\gamma/\sqrt{8}$. With $\gamma=1$, this is approximately $\Omega \simeq 0.35$. This coincides with the point where we observe the largest negativity (see Section~\ref{sec:semi-infinite}). This means that by canceling the coherent response from the system emission, it is possible to maximize its negativity. Having the largest negativity when the coherent reflectance $r$ is zero makes sense intuitively, since a coherent state has a positive Wigner function. 

In Appendix~\ref{app:incoherence} you can find a comprehensive discussion on what it means for the reconstructed state to be fully incoherent as defined by $\braket{\hat a}=0$.

%Mean intensity of the scattered field $\braket{a^\dagger a}$. Square of the mean field: $|\braket{a}|^2=\braket{\ahat}\braket{a^\dagger}$

%\textcolor{red}{PLOT INCOHERENT POWER AND CRAP}
%#############################################################################################################################################

\subsection{Purity}\label{sec:purity}
%Single-photon states are well known for their usefulness in quantum information and quantum computation schemes \textcolor{red}{[refs]}.

%If Wigner negativity is a resource, then not all single photon states

The states that are created in our setup are generally multiphoton states. For states of this kind, it is difficult to determine under which conditions their corresponding Wigner functions become negative. To understand the origin of negativity, we can restrict to the two-dimensional Fock space spanned by $\{\ket 0,\,\ket 1\}$ which simplifies the analysis. For states in this subspace, it is clear that the nonclassicality is due to the single-photon contribution. Nevertheless, it should be noted that the negativity of a state does not only depend on the populations but also on the coherences, i.e. the off-diagonal elements in the density matrix. Here we show that the amount of coherence determines the photon populations required for the state to be Wigner-negative, and hence the purity of the state strongly influences the negativity. 

\begin{figure}[hbt!]
    \centering
    % exjobb_program/integrate_negativity_more.py
    \includegraphics[width=\columnwidth]{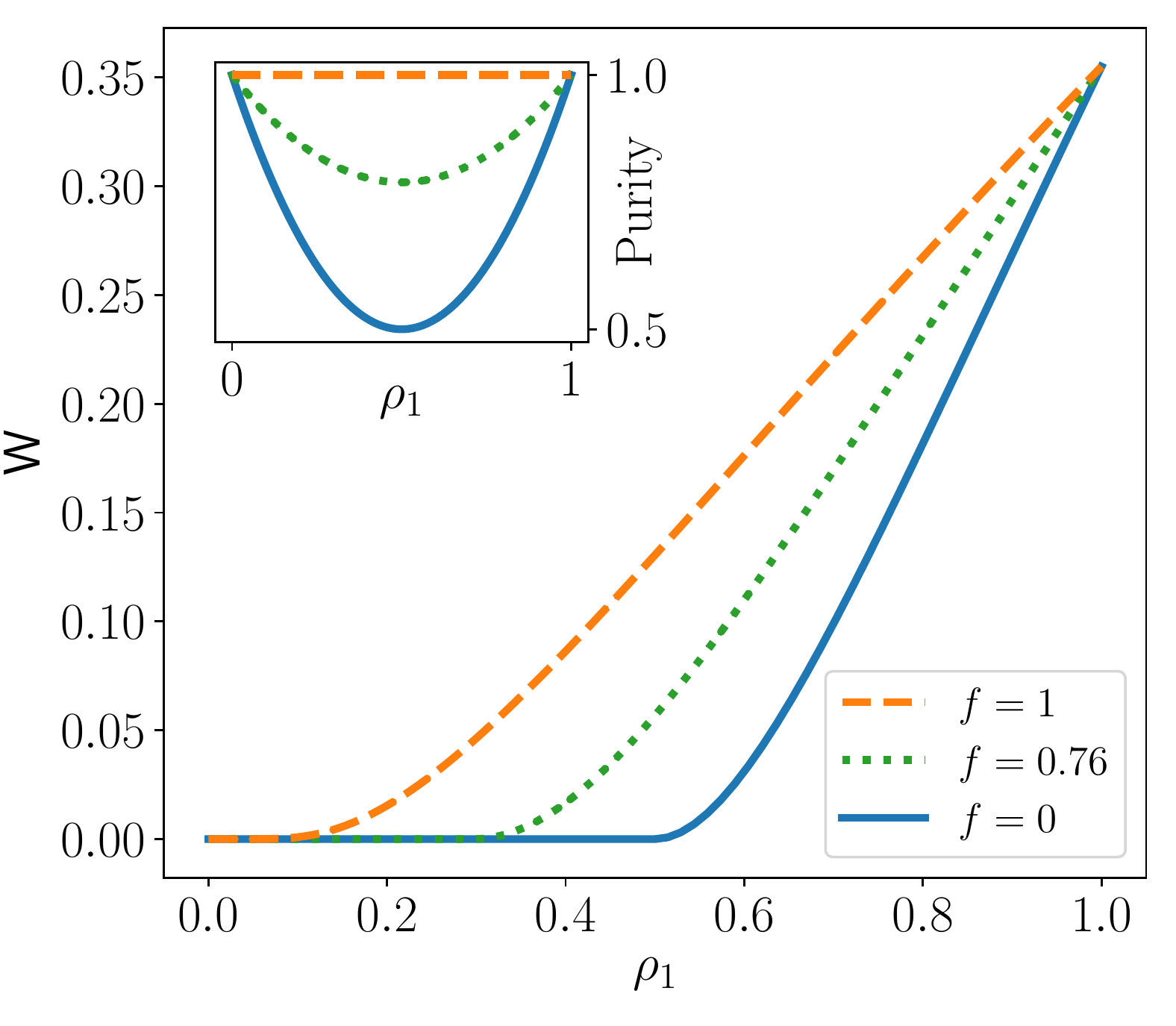}
    \caption{Integrated negativity of a pure state (dashed line) and two different statistical mixtures of $\ket 0$ and $\ket 1$ (dotted and solid line) as a function of the single-photon probability $\rho_1$. The negativities only coincide when the state is vacuum or single-photon with purity 1. Larger values of $f$ requires a smaller single-photon population in order for the state to be Wigner-negative. Insert: The purity of the same states as a function of the single-photon probability $\rho_1$. }
    \label{fig:purity_N}
\end{figure}

Both the populations and the coherences of a state determine the purity. A general state in the subspace $\{\ket 0,\,\ket 1\}$ is described by a density matrix of the form
\begin{equation}
\rho=
    \begin{pmatrix}
    \lvert\alpha\rvert^2 & f\alpha^*\beta \\
    f\alpha\beta^* & \lvert\beta\rvert^2
    \end{pmatrix},
\end{equation}
where the populations are $\lvert\alpha\rvert^2=\rho_0$ and $\lvert\beta\rvert^2=\rho_1$, and $\rho_0+\rho_1=1$. The parameter $f\in[0,1]$ modulates the coherence between the states $\ket 0$ and $\ket 1$. For $f=1$, we have a pure superposition of $\ket 0$ and $\ket 1$, while for $0 \le f <1$ we have a mixed state. For $f=1$, the purity of the state is not affected by the populations. However, this is not the case for $f<1$ where the purity varies with the photon content (see insert in Fig.~\ref{fig:purity_N}). The minimum purity of 0.5 is obtained by a maximally mixed state: $f=0$ and $\rho_0=\rho_1$. % The parameter $f$ sets a lower bound on the purity; for any state with $f>0$, the purity is always larger than 0.5. 

%The purity of this state is $\tr[\rho^2]=1-2(\rho_0-\rho_0^2)(1-f)$. 

As an example of how coherences affect the photon content needed to achieve Wigner negativity, we show the WLN for different values of $f$ in Fig.~\ref{fig:purity_N}. In the figure it can be seen that in order to get $\neg>0$ with the statistical mixture given by $f=0$, we are required to have $\rho_1>0.5$, which implies $\rho_1 >\rho_0$. This was already discussed in Section~\ref{sec:infinite_waveguide}. On the other hand, for a pure state there is no such restriction, and the Wigner function becomes negative already with a minuscule single-photon population. 
All other states with $0<f<1$ lie between the curves for the pure state and statistical mixture. 
%
%For a plot of purity as a function of time for different drive strengths, see Fig. 1 in the Supplemental material of~\cite{Quijandrna2018Dec}.

%The decrease in purity when adding a dephasing channel is central for explaining the subsequent reduction of negativity, as will be explained in the following section.

\begin{figure}[hbt!]
%ingrid/exjobb_program/NEWDATA2/plot_fock_dist_2chan_insert.py
        \includegraphics[width=0.85\columnwidth]{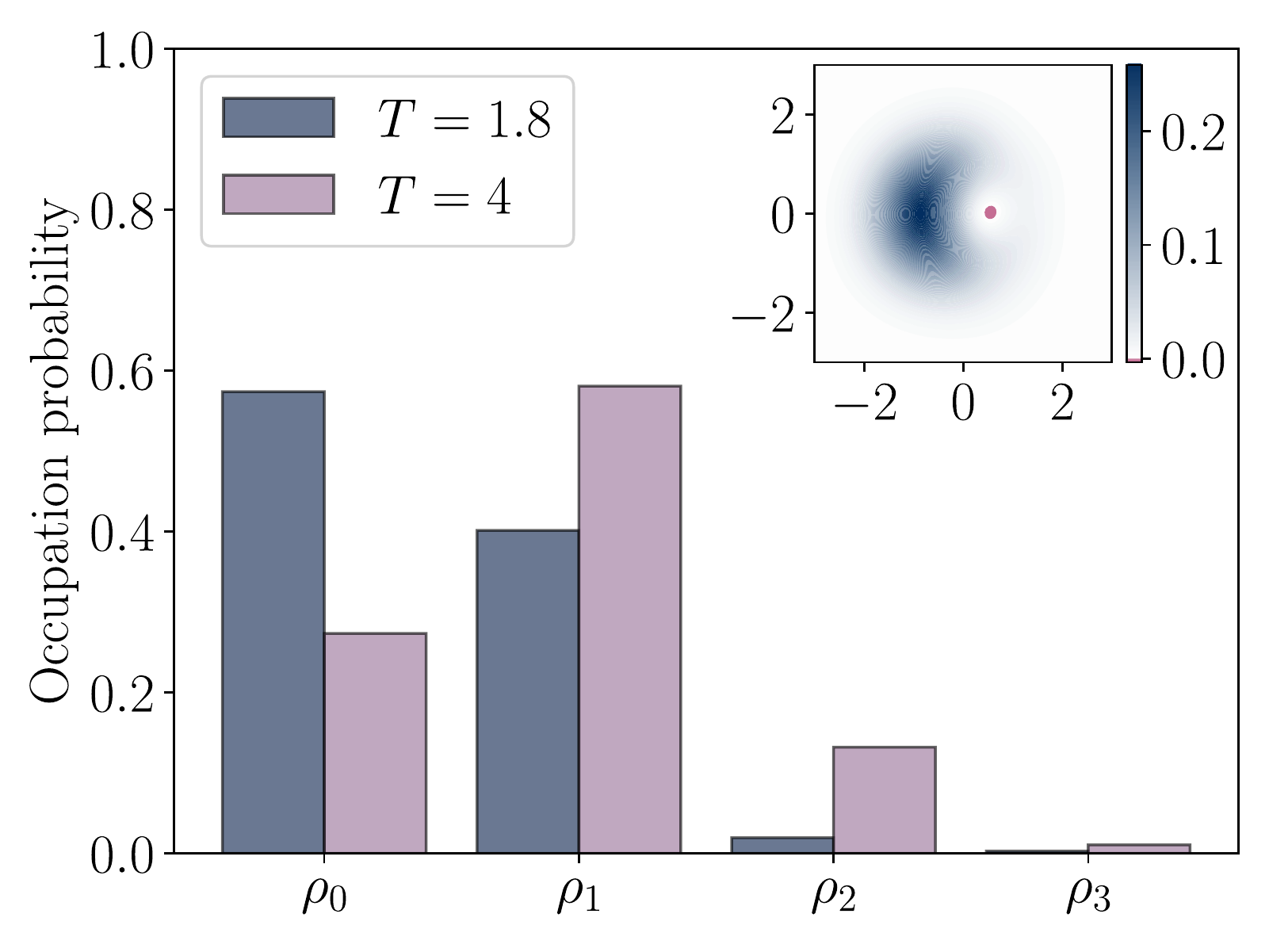}
        \label{res_2chan_pops}
    \caption{The diagonal elements of the density matrix for the reconstructed states with $\Omega=1/\sqrt{8}$ in the semi-infinite waveguide. $T=1.8$ is the integration time for which negativity is no longer negligible. Integration time $T=4$ gives the state with maximum observed negativity. The unit is $\gamma=1$. Inset: Wigner function for $T=1.8$. The Wigner function for $T=4$ is showed in Fig.~\ref{wigner_1chan}. }
    \label{pops}
\end{figure}

We can use this to understand our results. With the optimal drive strength $\Omega=1/\sqrt{8}$, consider $T=1.8$ which is the integration time for which Wigner negativity starts to become noticeable (in the sense that the WLN reaches a value over 0.0001). As can be seen in Fig.~\ref{pops}, the two-photon contribution is negligible for this state, which means our previous argument can be applied. For this state, the vacuum population is larger than the single-photon population which is $\rho_1=0.4$. From the reconstructed density matrix elements we can calculate $f=\lvert\rho_{01}\rvert/\sqrt{\rho_0 \rho_1}\approx 0.76$, which is consistent with observing negativity for $\rho_1<0.5$. However, for this value of $f$ negativity should start to be discernible already for $\rho_1=0.3$ according to the theoretical line in Fig.~\ref{fig:purity_N}. This discrepancy could be attributed to variations in the reconstructed density matrices (see Appendix~\ref{app:precision}).

In Fig.~\ref{pops} we also plot the populations of the most Wigner-negative state we observed, obtained with the integration time $T=4$. It is a multiphoton state; the single-photon population is the dominant one, but there is also a non-negligible two-photon population, demonstrating that this setup can provide Wigner-negative states outside of the $\{\ket 0,\,\ket 1\}$ subspace.

Purity is central for understanding the reduction in negativity that occurs when a dephasing channel is added, as will be explained in the following section.

%%%%%%%!!!!!!!!!!!!!!!!!!!!!!!!!!!!!!!!

%In particular, the Hudson theorem guarantees that the sole pure states with a positive Wigner function are Gaussian ones

%\url{https://journals.aps.org/pra/abstract/10.1103/PhysRevA.91.042309} mixed states can be more nonclassical than pure states, for a number of nonclassicality measures. but the opposite for others." It is, however, a ‘relativist’ approach in the sense that if a state is less nonclassical than another state according to certain measure then it might be, more classical in the framework of another measure. Interestingly, these authors observe that the statistical mixture of states can be more nonclassical than their superpositions."

\subsection{Additional decoherence channels}\label{sec:dephasing}

Since we only observe Wigner negativity in the semi-infinite transmission like, we restrict to this setup from here on. So far, the only decay channel in this system has been decay into the monitored waveguide. However, in realistic systems there are always other, unwanted loss mechanisms that affect the emission into the monitored channel. Since decoherence is the process that transforms a quantum state into a classical state~\cite{Zurek2003May,Zurek2008Jan}, it can be expected that it will reduce the Wigner negativity. In this section we investigate the effects of two additional unmonitored decoherence channels, pure dephasing and nonradiative decay. We explore the effect these decay channels have on the negativity at the drive strength $\Omega=1/\sqrt{8}$ that is optimal without decoherence, and relate the results to decoherence rates in a realistic superconducting device.

\subsubsection{Pure dephasing}
\begin{figure}[hbt!]
    \centering
    \includegraphics[width=\columnwidth]{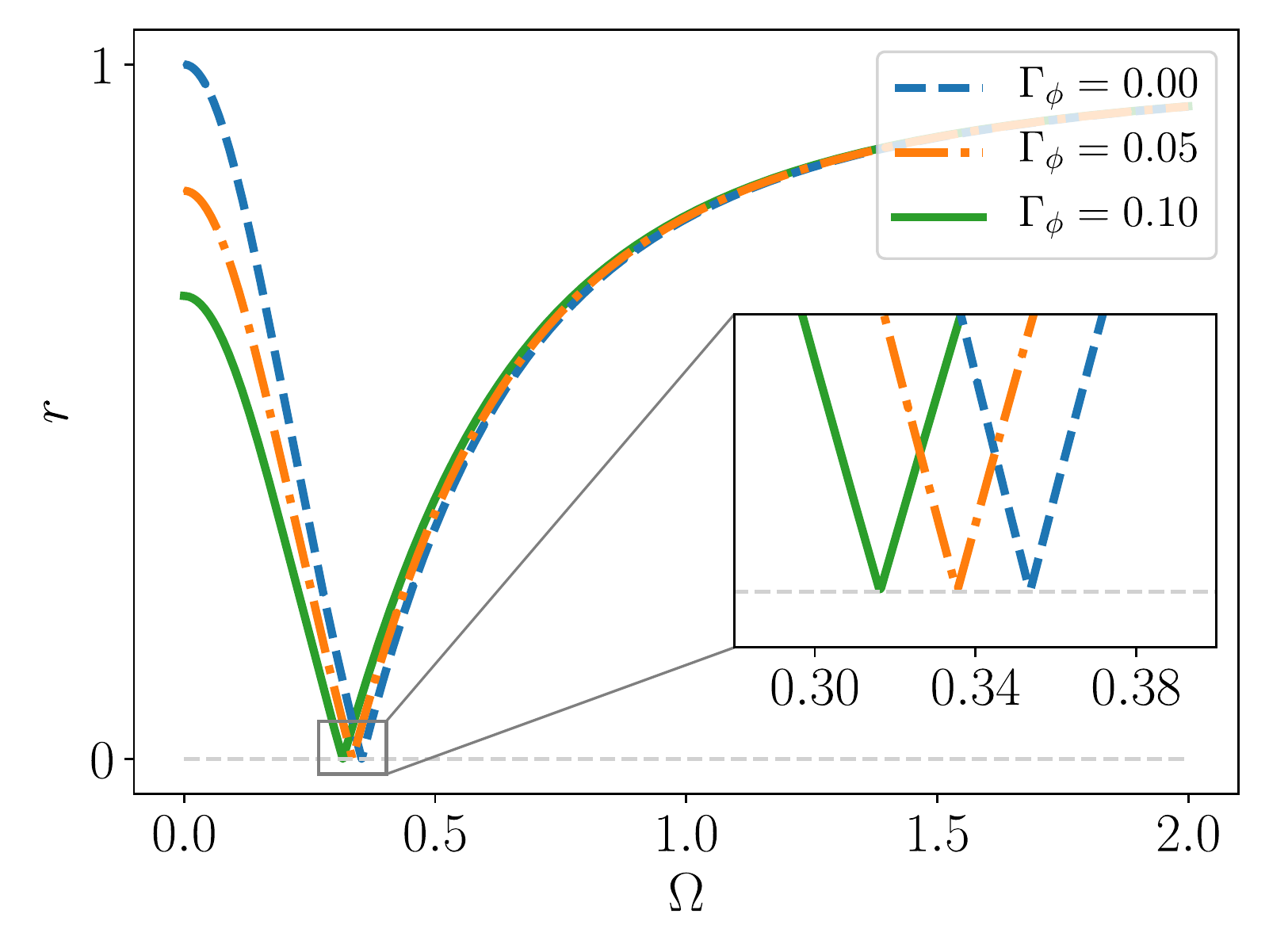}
    \caption{Coherent reflectance $r$ as a function of drive strength $\Omega$ for three different dephasing rates $\gammaphi$ (in units of $\gamma=1$), from Eq.~\eqref{eq:r_with_dephasing}. The zoom-in shows that the coherent reflectance is zero at different drive strengths for different dephasing rates.}
    % ~/exjobb_program/decoherence/pure_dephasing/coherent_reflectance.py
    \label{coherent_refl_plot}
\end{figure}

%\textcolor{red}{write that the plots are averages in  Figures~\ref{fig:only_dephasing,fig:nonrad,fig:rad_and_dephasing} and~\ref{purity}
%In figures~\ref{fig:only_dephasing,fig:nonrad,fig:rad_and_dephasing} and~\ref{purity}, every datapoint is an average over 50 simulation and reconstruction runs.
%}
We calculate the coherent reflectance with dephasing present in the system. Again solving for the steady-state, but now with the master equation
\begin{equation}
 \dot\rho=-\imi[H,\rho]+\gamma\mathcal{D}[\sigmaminus]\rho+\frac{\gammaphi}{2}\mathcal{D}[\sigma_z]\rho,
\end{equation}
where $\gammaphi$ is the pure dephasing rate, we get (setting \mbox{$\gamma=1$})
\begin{equation}
    \braket{\sigmaminus}=\frac{-2\Omega}{8\Omega^2+2\gammaphi+1}.
\end{equation}
Using the input-output relation~\eqref{inout} we get
\begin{equation}\label{eq:r_with_dephasing}
    r=\frac{\lvert\braket{a}\rvert}{\Omega}=\lvert1-\frac{2}{8\Omega^2+2\gammaphi+1}\rvert.
\end{equation}
Plotting Eq.~\eqref{eq:r_with_dephasing} for different dephasing rates in Fig.~\ref{coherent_refl_plot}, we see that dephasing shifts the incoherent point towards lower values of $\Omega$.
%
%¤¤¤¤¤¤¤¤¤¤¤¤¤¤¤¤¤¤¤¤¤¤¤¤¤¤¤¤¤¤¤¤¤¤¤¤¤¤¤¤¤¤¤4

\begin{figure}[hbt!]
    \centering
    \includegraphics[width=\columnwidth]{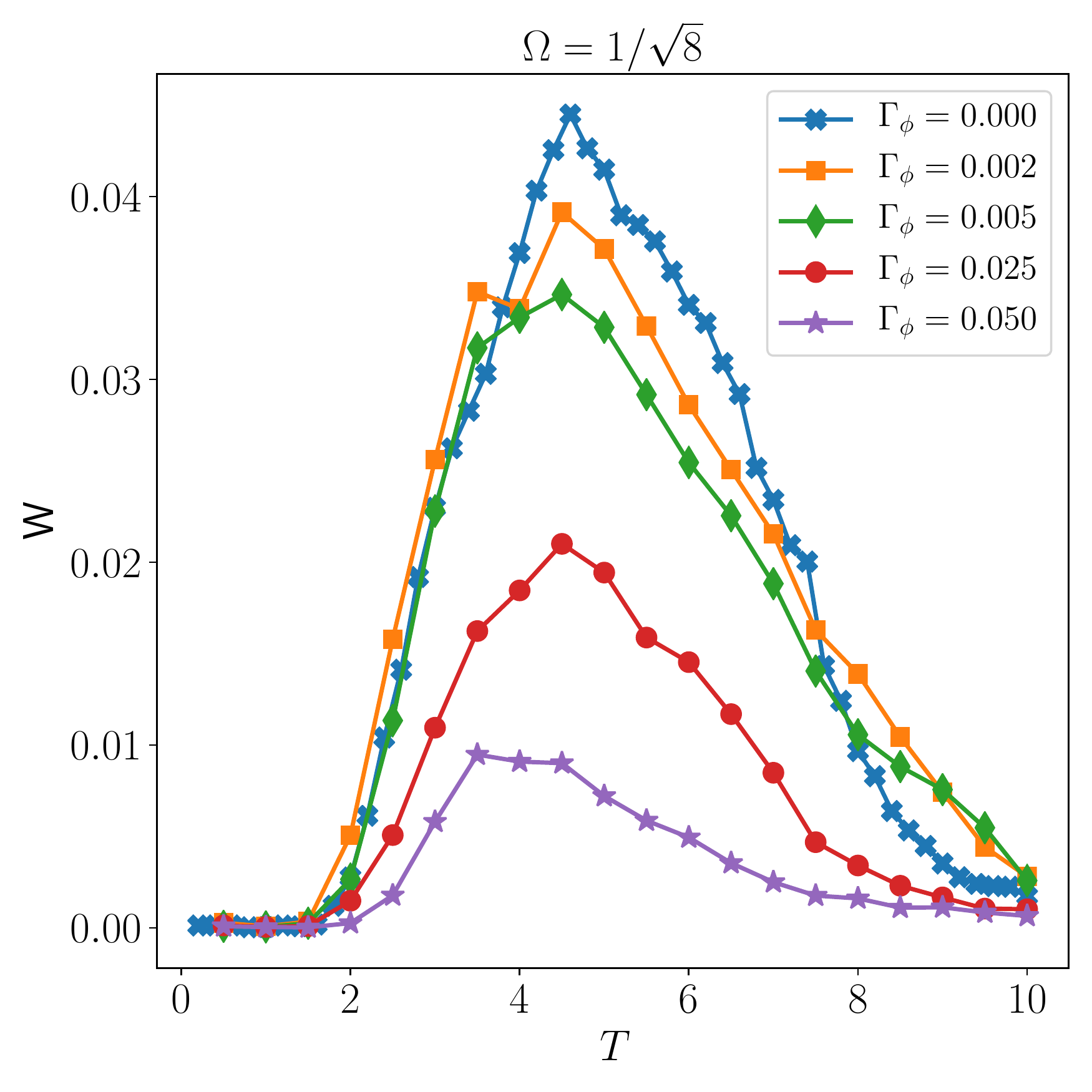}
    \caption{The WLN for different integration times and fixed $\Omega=1/\sqrt{8}$, with different dephasing rates $\gammaphi$, in units of $\gamma=1$. It is clear the WLN decreases with increasing dephasing rate. In this plot, each data point is an average over 50 reconstructions. See Appendix~\ref{app:precision} for a discussion of this.}
    % ~/exjobb_program/decoherence/pure_dephasing/plot_N.py
    \label{fig:only_dephasing}
\end{figure}

\begin{figure}[hbt!]
% plot from /exjobb_program/decoherence/pure_dephasing/T4/gamma_phi_0.0_new/plot_purity.py
% /exjobb_program/decoherence/pure_dephasing/T4/compare_purity.py
    \centering
    \includegraphics[width=\columnwidth]{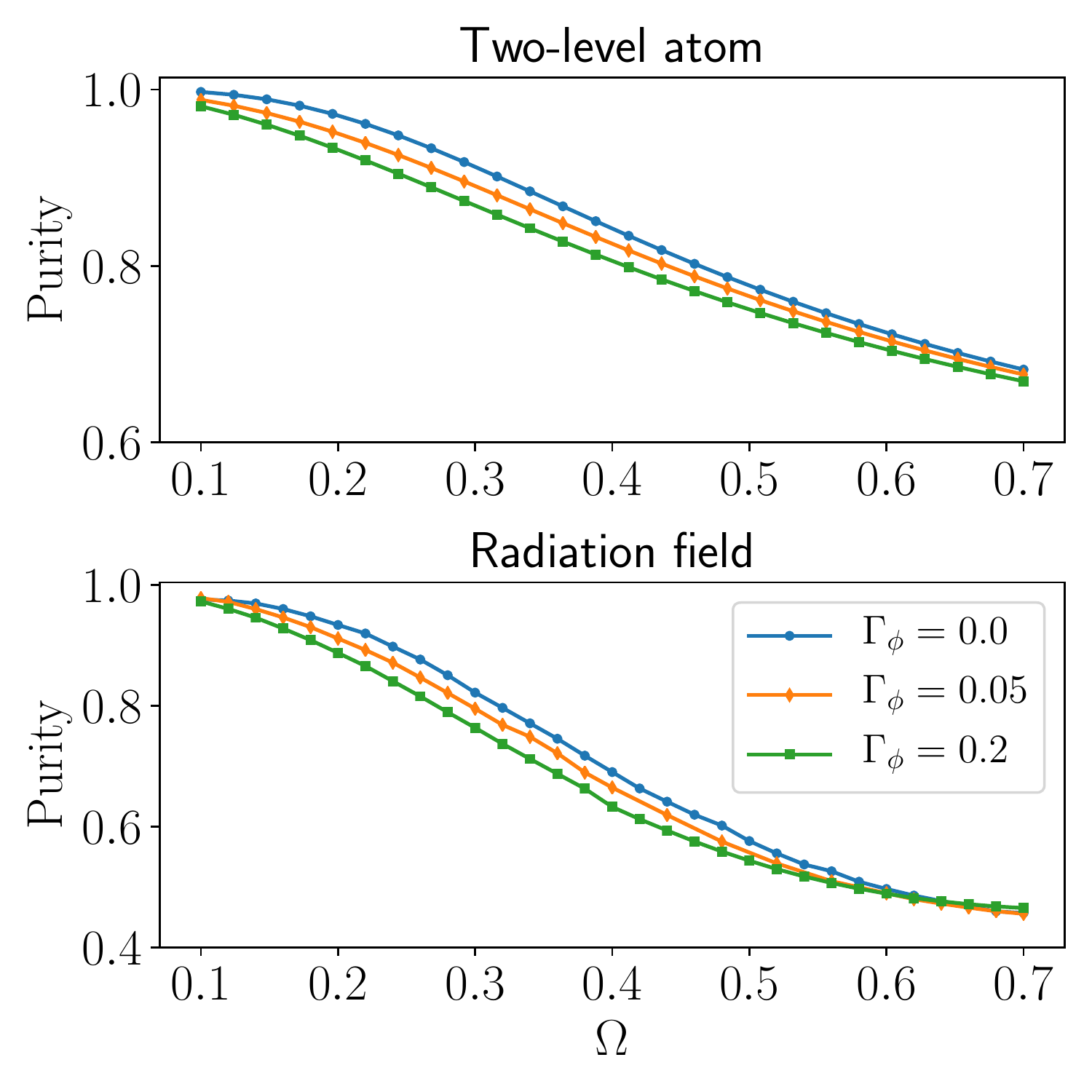}
    \caption{Simulation results for fixed $T=4$ (unit $\gamma=1$). The purity of the atom decreases with increased dephasing rate $\gammaphi$, with subsequent reduction of purity for the emitted radiation as well. The purity is also highly dependent on the drive strength. }
    \label{purity}
\end{figure}

In our numerical experiments, the effect of pure dephasing is reduced negativity for the emitted field, as can be seen in Fig.~\ref{fig:only_dephasing} for a fixed drive strength.
After the analysis in Section~\ref{sec:coherent_refl}, it would seem natural to assume that the reduction in negativity observed in Fig.~\ref{fig:only_dephasing} is due to decreased incoherent emission, since the incoherent point is shifted when dephasing is introduced. However, this is not the case, as we did not see improvement in negativity when driving at the new incoherent points (not shown). The reason for this is that the introduction of pure dephasing only affects the off-diagonal elements (coherences) in the atom density matrix, decreasing the purity. This translates into reduced coherences in the radiation field density matrix [see Fig.~\ref{purity}], while the populations are kept intact. Again looking at the example of the two-dimensional Fock subspace, loss of coherence means that a larger single-photon population is required for the state to be Wigner-negative, as showed in Section~\ref{sec:purity}. But with a fixed drive strength, the single-photon population is fixed, so the negativity of the state is diminished.

%!!!!!!!!!!!!!!!!!!!!!!!!!!!!!!!!!!!!!!!!!!!!!!!!!!!!!!!1
\subsubsection{Decoherence in a superconducting device}

\begin{figure}[hbt!]
    \centering
    \includegraphics[width=\columnwidth]{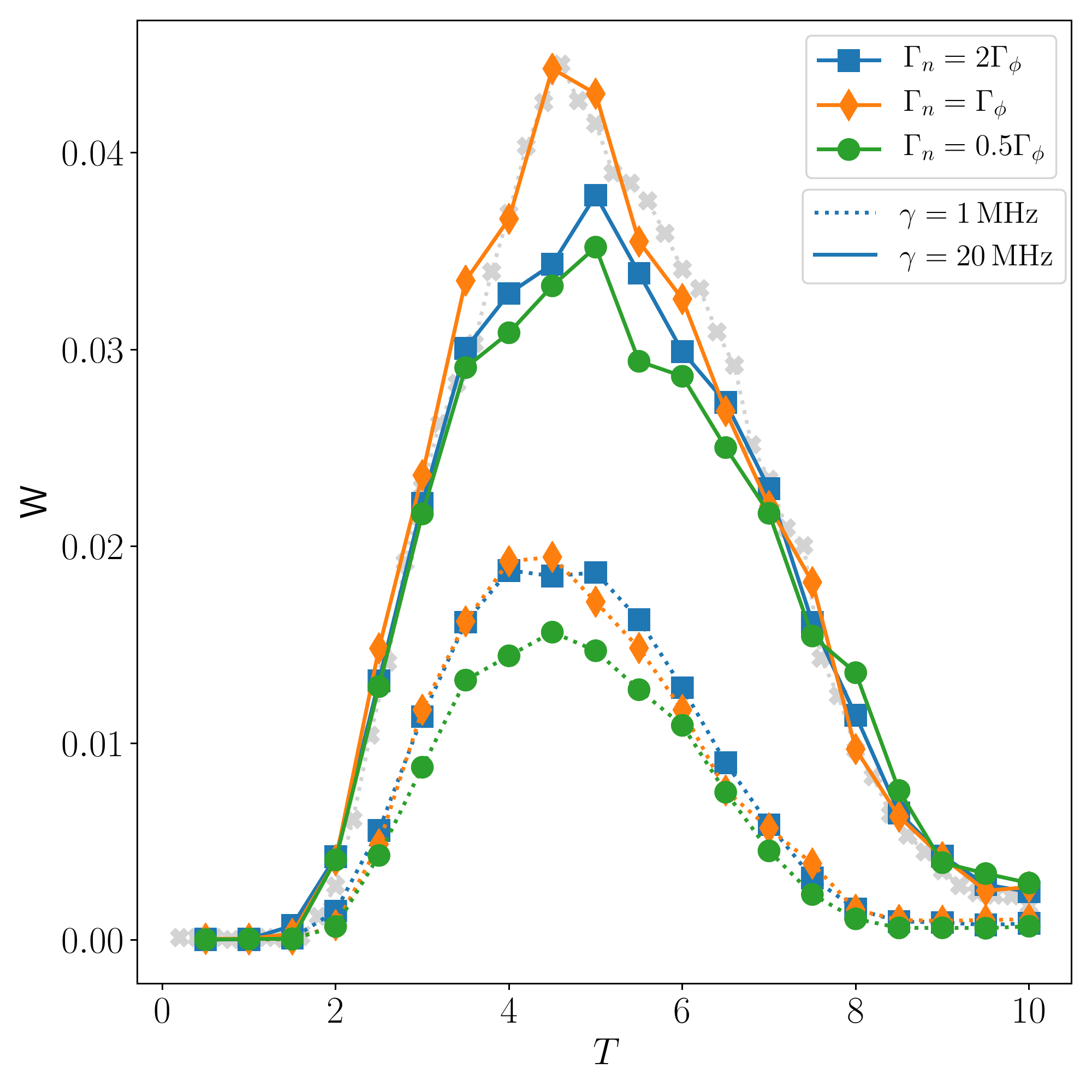}
    \caption{ The WLN for different integration times and fixed $\Omega=1/\sqrt{8}$ (in units of $\gamma$, see legend), with both nonradiative decay and dephasing according to the relation~\eqref{decay_rates}. The solid lines show radiative decay rate $\gamma=\SI{1}{MHz}$, and the dashed lines show $\gamma=\SI{20}{MHz}$. The gray dotted line shows the WLN without the additional decoherence. With $\gamma=\SI{20}{MHz}$, the decoherence decay rates are comparatively very small and thus barely affect the WLN. }
    % exjobb_program/decoherence/with_radiative_decay/plot_N_all.py
    \label{fig:rad_and_dephasing}
\end{figure}

% exjobb_program/dephasing/with_radiative_decay/plotN.py
To conclude, we will discuss a realistic implementation in a circuit-QED setup~\cite{Hoi2012Jun, Hoi2015Sep}. We consider a transmon qubit coupled to the end of a 1D transmission line. Along with pure dephasing, in an experimental setup there will also be losses due to nonradiative decay. Nonradiative losses correspond to an unmonitored decay channel, and as we saw in Section~\ref{sec:infinite_waveguide}, this results in a reduced Wigner negativity in the monitored decay channel. Additionally, as seen in the previous Section~\ref{sec:dephasing}, the effect of dephasing is also reduced negativity in the observed state. Nevertheless, if the dephasing rate $\gammaphi$ and decay rate $\gamman$ of the unmonitored channel are small compared to the radiative decay rate (\mbox{$\gammaphi, \, \gamman \ll \gamma$}), Wigner negativity can still be preserved. Fortunately, in a superconducting device, the ratio of the dephasing and nonradiative decay rates to the radiative decay rate can be very small~\cite{Yong2019}. We consider a sample device with measured decay rates  
\begin{equation}\label{decay_rates}
    \gamman+2\Gamma_\phi = \SI{89}{kHz},
\end{equation}
and radiative decay rate $\gamma=\SI{1}{MHz}$. However, devices can be manufactured with a radiative decay rate up to \SI{20}{MHz} while keeping the relation~\eqref{decay_rates}; this would result in barely any reduction of negativity compared to the ideal case with no unwanted decay channels. Figure~\ref{fig:rad_and_dephasing} shows the WLN for different rates of both these nonradiative decay processes, constrained to satisfy~\eqref{decay_rates}. Even for the smaller value of $\gamma$, some negativity is still present. 
This suggests that Wigner-negative states from one-dimensional resonance fluorescence can possibly be generated in an experimental circuit-QED setup.

%#############################################################################################################################################

%\subsection{Coherent displacement}
%\input{results/coherent_disp.tex}

%#############################################################################################################################################

%\subsection{Squeezing}
%\input{results/squeezing.tex}

%\subsection{Addition of noise}

%\subsection{Gaussianity}
%\input{results/gaussianity.tex}

%\subsection{Integrated negativity's sensitivity to changes in $\rho$, coherences, numerical errors(?)}\label{sec:sensitivity}
%\input{results/sensitivity.tex}

%\subsection{Entanglement}

%%%%%%%%%%%%%%%%%%%%%%%%%%%%%%%%%%%%%%%%%%%%%%%%%%%%%%%%%%%%%%%%%%%%%%%%%%%%%%%%%%%%%%%%%%%%%%%%%%%%%%%%%%%%%%%%%%%%%%%%%%%%%%%%%%%%%%%%%%%%%%%%%%%%%

\section{Conclusions}\label{conclusion}
% Move 1: Background info (research purpose, theory, method)
% Move 2: Summarizing key results
% Move 3: Commenting on key results (making claims, explaining results, comparing with previous works, etc)
% Move 4: Stating limitations
% Move 5: Recommendations for future implementations or future research

%\textcolor{blue}{Note: Mostly moves 1 and 2. Move 3 will appear in the Results section, which will also contain discussion.}

In this work, we have numerically studied the steady-state resonance fluorescence from a resonantly driven two-level system in a one-dimensional waveguide, with the purpose of generating quantum states that are nonclassical in the sense that they have a negative Wigner function.

The quantum state of the two-level system was evolved in time by solving the stochastic master equation for homodyne detection. After steady-state had been reached, the homodyne measurement results of the output resonance fluorescence were recorded for a time $T$, and maximum-likelihood estimation was used to reconstruct the density matrix of the output field. From the density matrix, the Wigner function of the resonance fluorescence was calculated. Recently, an alternative formalism for the study of filtered propagating modes was introduced~\cite{Kiilerich2019Feb}. We have confirmed our results with this method.

Because quantum states with a negative Wigner function have been identified as a necessary resource to achieve a quantum speedup for continuous variable quantum computing~\citep{Mari2012Dec,Veitch2013Jan}, 
%we investigated which combinations of parameters $(\Omega,T)$ can produce output radiation that belongs to this class of states. For our two setups, we found that producing Wigner-negative states is only possible with the semi-infinite waveguide. 
we investigated in which parameter regimes Wigner-negative states can be generated. For our two setups, we found that producing states belonging to this class is only possible with the semi-infinite waveguide. In particular, maximum negativity is achieved when the coherent response from the system is entirely suppressed. On the other hand, with an infinite waveguide we only observed positive Wigner functions, due to the contribution of vacuum that appears when only monitoring one side of the transmission line.

We showed that the purity of the state affects the negativity. A state with high purity can exhibit Wigner-negativity even when having a large vacuum contribution that would render a mixed state positive. Furthermore, we examined the effects of decoherence---specifically, pure dephasing and nonradiative decay. While decoherence in general reduces the negativity, we found that if the decoherence rates are much smaller than the radiative decay rate, which is realistic for superconducting devices, the impact on negativity can be negligible. 

%---a detector efficiency below \textcolor{red}{insert result}, corresponding to \textcolor{red}{insert number} of thermal photons at \textcolor{red}{insert frequency} was shown to completely destroy any negativity. 

This setup is appealing due to its simplicity, with resonance fluorescence having already been observed experimentally with trapped ions~\cite{Hoffges1998Jan}, superconducting circuits~\cite{Astafiev840} and quantum dots~\cite{Muller2007Nov}. However, while the setup can generate Wigner-negative states, it is not yet clear how to utilize them for quantum information processing. Together with Gaussian operations which are relatively easy to implement, an additional non-Gaussian operation is required for universal quantum computation~\cite{Lloyd1999Feb}. This type of operation can be created by non-Gaussian, or Wigner-negative, states via gate teleportation. There are known protocols to implement the lowest order non-Gaussian gate, the cubic phase gate, by producing resource states such as the cubic phase state~\cite{Gottesman2001Jun,Sabapathy2018Jun,Ghose2007Apr}. The challenge remains to design a protocol that can generate a useful non-Gaussian operation from our type of state. Another possibility is that the states could be used for resource concentration, where less resourceful non-Gaussian states are used to produce more resourceful outputs by only Gaussian operations~\cite{Albarelli2018Nov}.
For further work, it would also be of interest to investigate whether it is possible to optimize the temporal mode filter function to maximize the Wigner negativity.

%it could be interesting to investigate if and how the introduction of a distance between the atom and the mirror affects the Wigner negativity. 

%\textcolor{red}{future:  distance between atom and mirror, optimize mode function \url{https://journals.aps.org/pra/abstract/10.1103/PhysRevA.99.033832}, entanglement with other modes}

%https://journals.aps.org/pra/pdf/10.1103/PhysRevA.98.032316

%There are Gaussian protocols that can take less resourceful non-Gaussian states and produce more resourceful outputs, known as resource concentration. 

%The challenge then is to design a particular circuit that can generate a useful nonlinear (non-Gaussian) operation, for example a cubic phase gate on a single mode.

%Protocols have been demonstrated where, e.g., the cubic  phase  gate,  which  allows  to  promote  the  Gaussian set  of  gates  to  a  universal  set  [9],  can  be  obtained  by using input non-Gaussian ancillary states together with Gaussian operations and measurements [32, 54]. Therefore, we have verified that this simple setup suffices to generate the class of states necessary for universal quantum computing beyond the scope of classical computers

%have been identified as a necessary resource to achieve a quantum speedup for continuous variable quantum computation~\citep{Mari2012Dec,Veitch2012Nov,Veitch2013Jan}.

\begin{acknowledgments}
We thank Claude Fabre for discussions about the temporal mode function.
FQ and GJ acknowledge financial support from the Knut and Alice Wallenberg Foundation.
IS acknowledges support from Chalmers Excellence Initiative Nano.

% we would like to thank <name> dor fruitful discussions
% we acknowledge helpful discussions with <name
% we thank <name> for 
% we wish to thank
% we are grateful to <name>
% useful discussions
\end{acknowledgments}

\appendix

\section{Wigner function in the Fock basis}\label{sec:app1}

For a state 
\begin{equation}
    \rho = \sum_{nm} \rho_{nm} \ketbra{n}{m}
\end{equation}
where $\ket n$ is a Fock basis state, the corresponding Wigner function can be expressed as
\begin{equation}
  W(x,p) = \sum_{mn} \rho_{mn} W_{mn}(x,p),
\end{equation}
with the matrix elements $W_{mn}$ given by
\begin{widetext}
\begin{equation}
\begin{cases}
W_{mn} = \frac{1}{\pi}  \expup{-(x^2+p^2)}  (-1)^m \sqrt{2^{n-m}\frac{m!}{n!}}(x-\imi p)^{n-m} L_m^{n-m}(2x^2+2p^2),\quad n\geq m ,\\
W_{mn} = \frac{1}{\pi}  \expup{-(x^2+p^2)} (-1)^n \sqrt{2^{m-n}\frac{n!}{m!}}(x+ip)^{m-n} L_n^{m-n}(2x^2+2p^2), \quad n<m.
\end{cases}
\end{equation}
\end{widetext}

%%%%%%%%%%%%%%%%%%%%%%%%%%%%%%%%%%%%%%%%%%%%%%%%%5

\section{Coherence, incoherence and phase space}\label{app:incoherence}

We would like to clear up  confusion than could potentially arise from using the descriptions "coherent" and "incoherent" in different contexts. There are two properties that are commonly referred to as "quantum coherence". For general quantum states, what is called \emph{quantum coherence} is phase information encoded by the off-diagonal elements of the density matrix. In the field of quantum optics, the concept of coherence introduced by Glauber~\cite{Glauber1963Jun,Glauber1963Sep} is related to the classical possibility of producing interference fringes when two fields are superimposed~\cite{Glauber2007Jan}. There are different orders of this \emph{quantum optical coherence}: first-order coherence, second-order coherence, and so on. In particular, a state is $n$th-order coherent when its $n$th-order correlation function factorizes. Quantum coherence of both kinds are linked to the possibility of interference, but of probability amplitudes instead of field amplitudes as in the classical setting. %There are different measures for the two types of coherence. For instance, the degree of first-order coherence determined by the value of the normalized first-order correlation, the $g^{(1)}$-function: $g^{(1)}=\braket{a^\dagger a}/\lvert \braket{a}\rvert^2$, with $\lvert g^{(1)}\rvert=1$ when the field is first-order coherent. In contrast, quantification of quantum coherence between different quantum states is not so straight-forward, and several different measures have been proposed (see~\cite{Jin2018Jun} and references therein).
We have used both definitions of coherence: in Section~\ref{sec:coherent_refl} we used first-order coherence of the resonance fluorescence, and in Section~\ref{sec:purity} and forward we used quantum coherence between Fock states.

%A quantum state which is a statistical mixture, sometimes also called incoherent mixture or incoherent superposition, has a diagonal density matrix, whereas a coherent superposition has off-diagonal elements as well. Quantum states can be coherent in one sense or the other, neither, or both\cite{ballentine1998quantum}. For example, a single-mode field is always first-order coherent, whether it is in a pure or mixed state~\cite{Glauber2007Jan,ballentine1998quantum}. 

%This state is incoherent in the sense that there is no phase information. 

For a state carrying no phase information, for instance a Fock state, the Wigner function is symmetric in phase space. With our setup at the incoherent drive point, all of the emitted radiation is incoherent in the sense that there is no phase relation between the average emitted field and the drive. Despite this, the observed states are not symmetric in phase space (as seen in Fig.~\ref{wigner_1chan}). This is because while there is no phase relation between the average field and the drive, there can still be correlations between subsequently emitted photons. The second-order correlation function $g^{(2)}(\tau)$ describes correlations between photons. When $g^{(2)}(\tau)=1$ the photon emissions are uncorrelated. While the $g^{(2)}$-function for resonance fluorescence stabilizes at 1 for some $\tau^*>0$, resonance fluorescence exhibits antibunching of photons, as indicated by $g^{(2)}(0)=0$~\cite{Kimble1977Sep, Walls1994}. Consequently, there are correlations between photons emitted with a time difference shorter than $\tau^*$, and thus it is natural that the final state measured over a time $T\lesssim\tau^*$ contains coherences. 

\subsection{Coherent displacement}

For the results so far, we have simulated detection of the atomic emission only, i.e. ignored the reflected drive field since it only shifts the Wigner function in phase space without affecting the negativity (shown in~\cite{Quijandrna2018Dec}). In this section we include the drive, and also allow it to be complex by adding a phase: $\Omega=\lvert \Omega \rvert \expup{\imi\varphi}$, such that with $\gamma=1$ the Hamiltonian in Eq.~\eqref{drive_hamiltonian} becomes $\hat H =-\imi(\Omega\sigmaplus + \Omega^*\sigmaminus)$.

While the maximally negative state is not symmetric in phase space, it is centered around the origin. This is because $\braket{\ahat_f}=0$ at the incoherent drive point. In fact, all states are displaced from the origin by %
\begin{equation}\label{ahat}
   \Delta= \sqrt{2}\braket{\ahat_f}=\sqrt{2T}\left(\lvert\Omega\rvert-\frac{2\lvert\Omega\rvert}{1+8\lvert\Omega\rvert^2}\right),
\end{equation}
where the $\sqrt{2}$ is due to the normalization chosen for the quadratures in~\eqref{bothquads}, and the factor of $\sqrt T$ comes from integrating~\eqref{propermode} over our choice of filter function~\eqref{eq:boxcar}. 
\begin{figure}[hbt!]
% exjobb_program/sensitivity/analysis.py
    \centering
    \subfloat[Reconstructed state for $T=100$ and $\Omega=0.05$. It has fidelity 0.95 with a coherent state with the same number of photons.]{
        \includegraphics[width=\columnwidth]{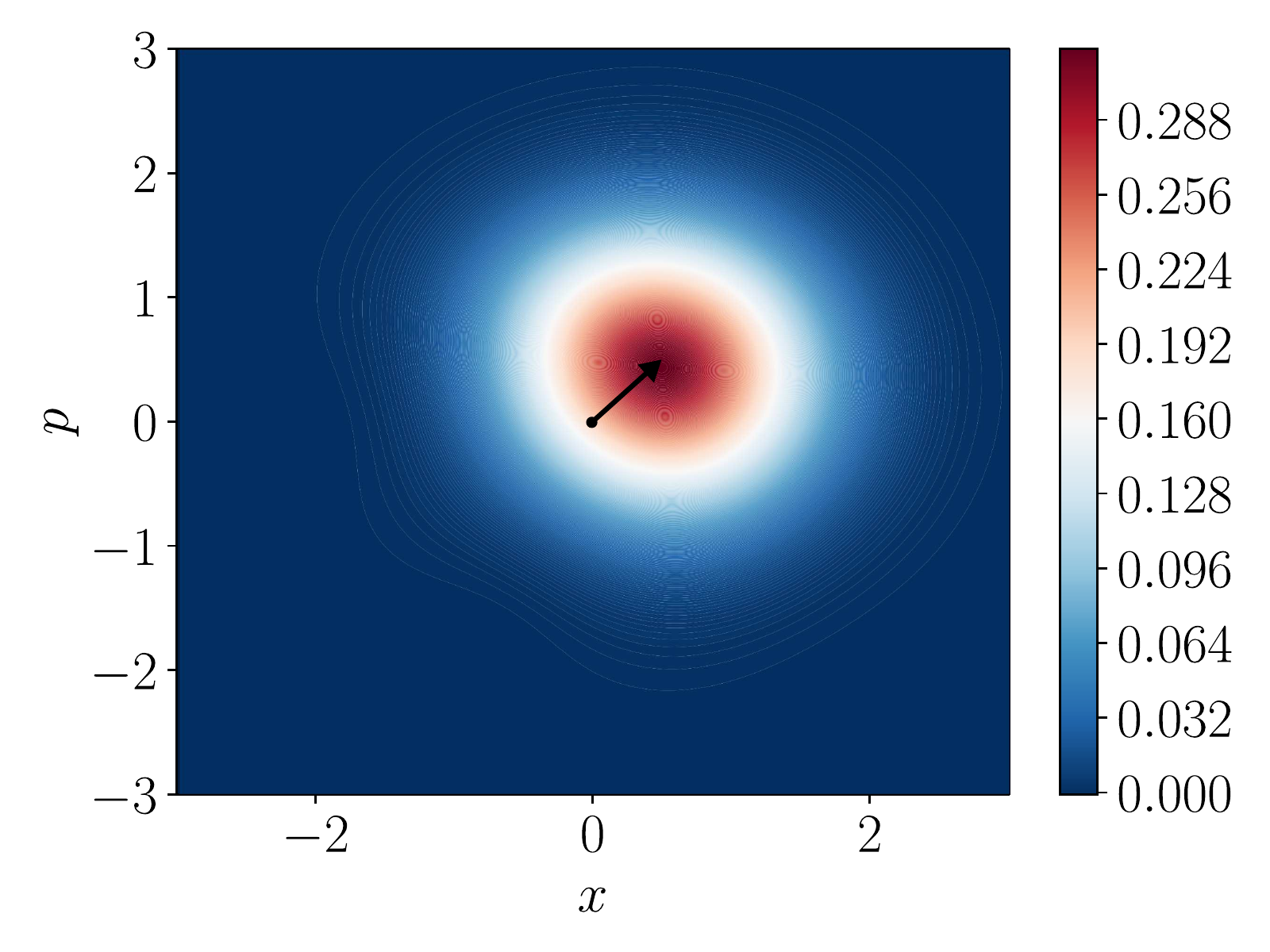}
                \label{fig:displacement_coh}
        }
        %\caption{Reconstructed state for $T=100$ and $\Omega=0.05$. It has fidelity 0.95 with a coherent state with the same number of photons.}
    
    \subfloat[Reconstructed state for $T=30$ and $\Omega=0.5$. It has fidelity 0.18 with a coherent state with the same number of photons.]{
        % exjobb_program/sensitivity/ corr_purity_negativity.py
        \includegraphics[width=\columnwidth]{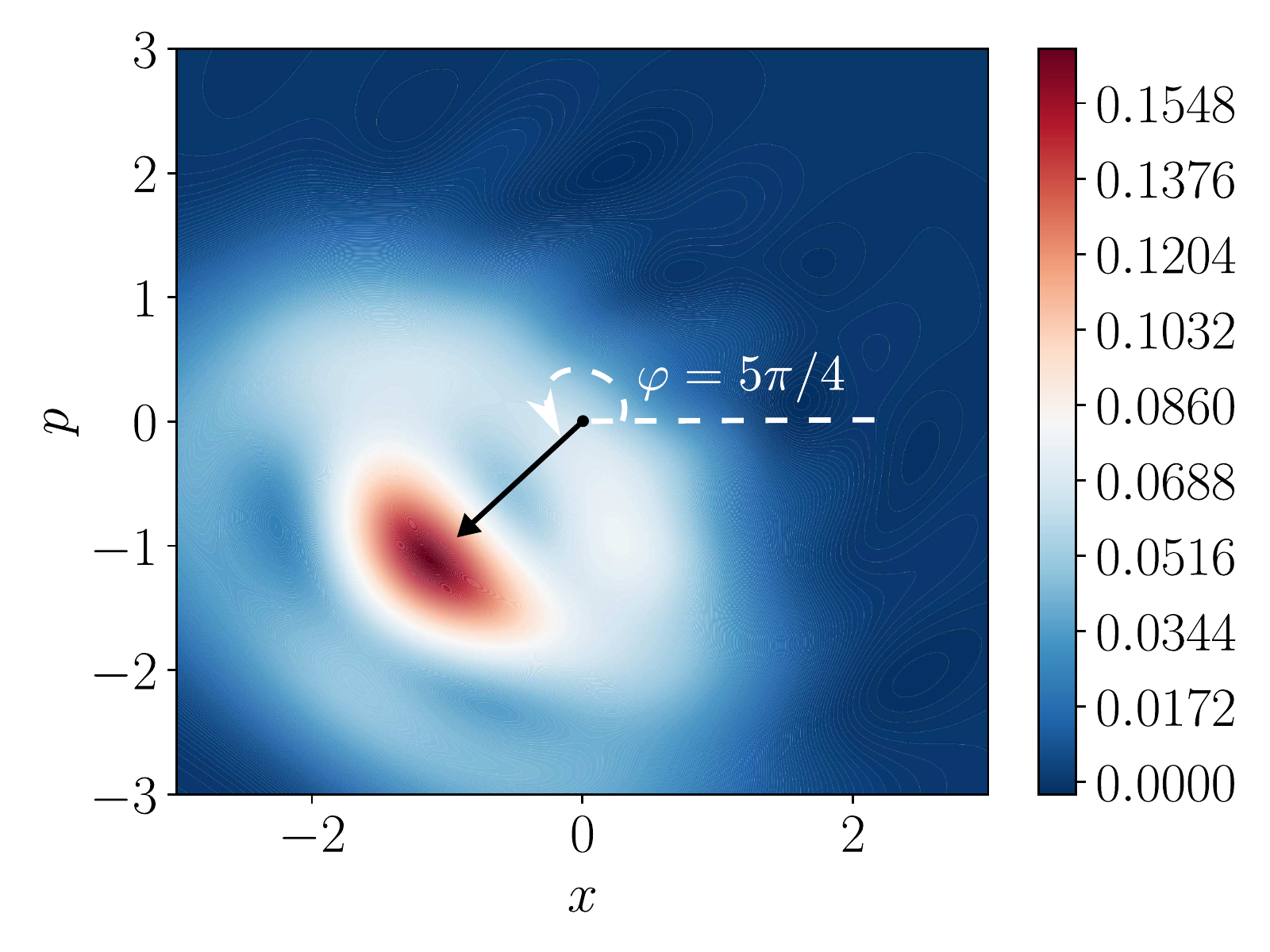}
             \label{fig:displacement_incoh}}
        %\caption{Reconstructed state for $T=30$ and $\Omega=0.5$. It has fidelity 0.18 with a coherent state with the same number of photons. }
   
    \caption{States displaced from the origin on a line in phase space given by the phase $\varphi=5\pi/4$ of the driving field, in the direction given by the sign of $\braket{\ahat_f}$ in~\eqref{ahat}. The lengths of the arrows are given by $\Delta$.}
    \label{fig:displacement}
\end{figure}
In the very weak driving regime, the atom scatters essentially all of the incoming field coherently (see Fig.~\ref{coherent_refl_plot}). A coherent state $\ket{\alpha}$ has $\lvert \alpha \rvert^2$ photons and is displaced from the origin by $\sqrt{2}\lvert \alpha \rvert$. The number of photons sent by the drive field to be reflected during a time $T$ is $T\Omega^2$. When reflected coherently, the observed state is expected to be displaced by $\sqrt{2T}\lvert\Omega\rvert$. This is confirmed by simulations, and also by taking the limit of $\Omega \ll 1$ in Eq.~\eqref{ahat}, which gives $\Delta =-\sqrt{2T}\Omega$. For a real drive $\Omega$ which we used before, the state is shifted in the negative $x$-direction. The direction is determined by the phase $\varphi$ of the driving field. In Figs.~\ref{fig:displacement}, the phase is $\varphi=5\pi/4$. Figure~\ref{fig:displacement_coh} shows the displacement of a reflected approximately coherent state, and Fig.~\ref{fig:displacement_incoh} shows the same type of displacement for an example state that is not coherent, visualizing that the displacement of any type of state is given by Eq.~\eqref{ahat}.

\section{Precision of the maximum-likelihood reconstruction}\label{app:precision}

The fact that outcomes of quantum mechanical measurements are inherently random, combined with only having a finite number of measurements, induce a statistical uncertainty in the result of the maximum-likelihood state reconstruction. Quantifying the uncertainty for quantum tomography is not straightforward and there are many theoretical  approaches~\cite{Blume-Kohout2012Feb,Audenaert2009Feb,Li2016Dec,Oh2019Feb,Usami2003Aug,Faist2016Jul}. In practice, the simplest way is to use the \emph{bootstrapping method}: generating an ensemble of simulation results with the same parameter settings and report the variation of the reconstructed density matrices~\cite{Suess2017Sep}. Here we do this for the maximally negative state at the incoherent point, reconstructed 80 times. We also compare the spread in the results for 500 vs. 1000 simulation trajectories, and 20 vs. 40 tomography angles.

\begin{figure}[hbt!]
    \centering
    % exjobb_program/sensitivity/boxplot_fidelities.py
    \includegraphics[width=\columnwidth]{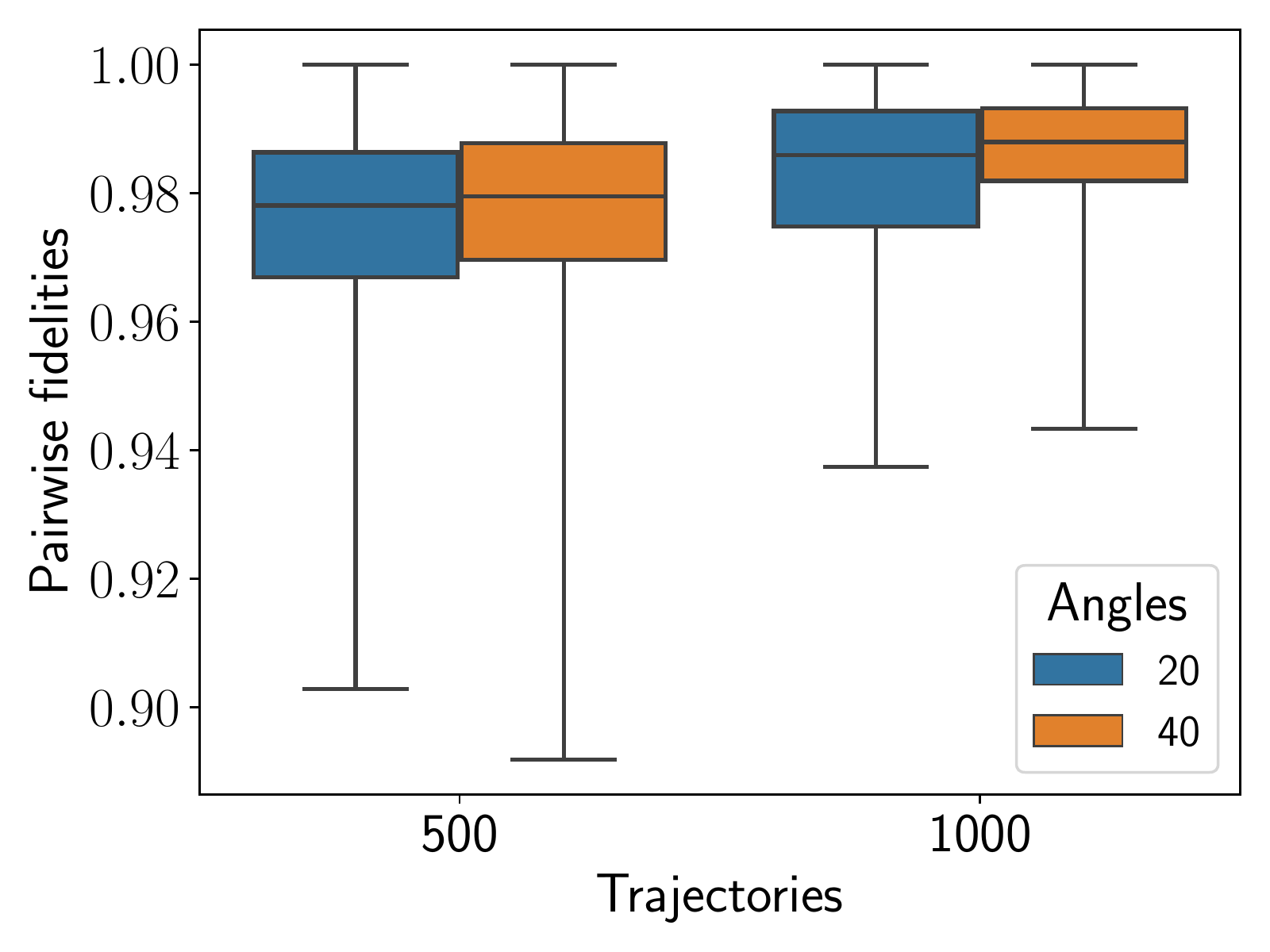}
    \caption{Pairwise fidelitites between 80 reconstructions from the atom in the semi-infinite waveguide at the incoherent point. The box shows the quartiles of the dataset while the whiskers extend to show the rest of the distribution. The horizontal line indicates the median. The histogram of the leftmost box (500 trajectories and 20 angles) can be seen in Fig.~\ref{fig:pair_fidelity}. }
    \label{fig:box_fid}
\end{figure}

Figure~\ref{fig:box_fid} shows a boxplot of the pairwise fidelities between all 80 states. Using 1000 trajectories in the simulations clearly reduce the variance compared to 500 trajectories. There is however no clear difference between using 20 and 40 angles for the tomography.

\begin{figure}[hbt!]
    \centering
    %   % exjobb_program/sensitivity/pairwise_fidelities.py
    \includegraphics[width=\columnwidth]{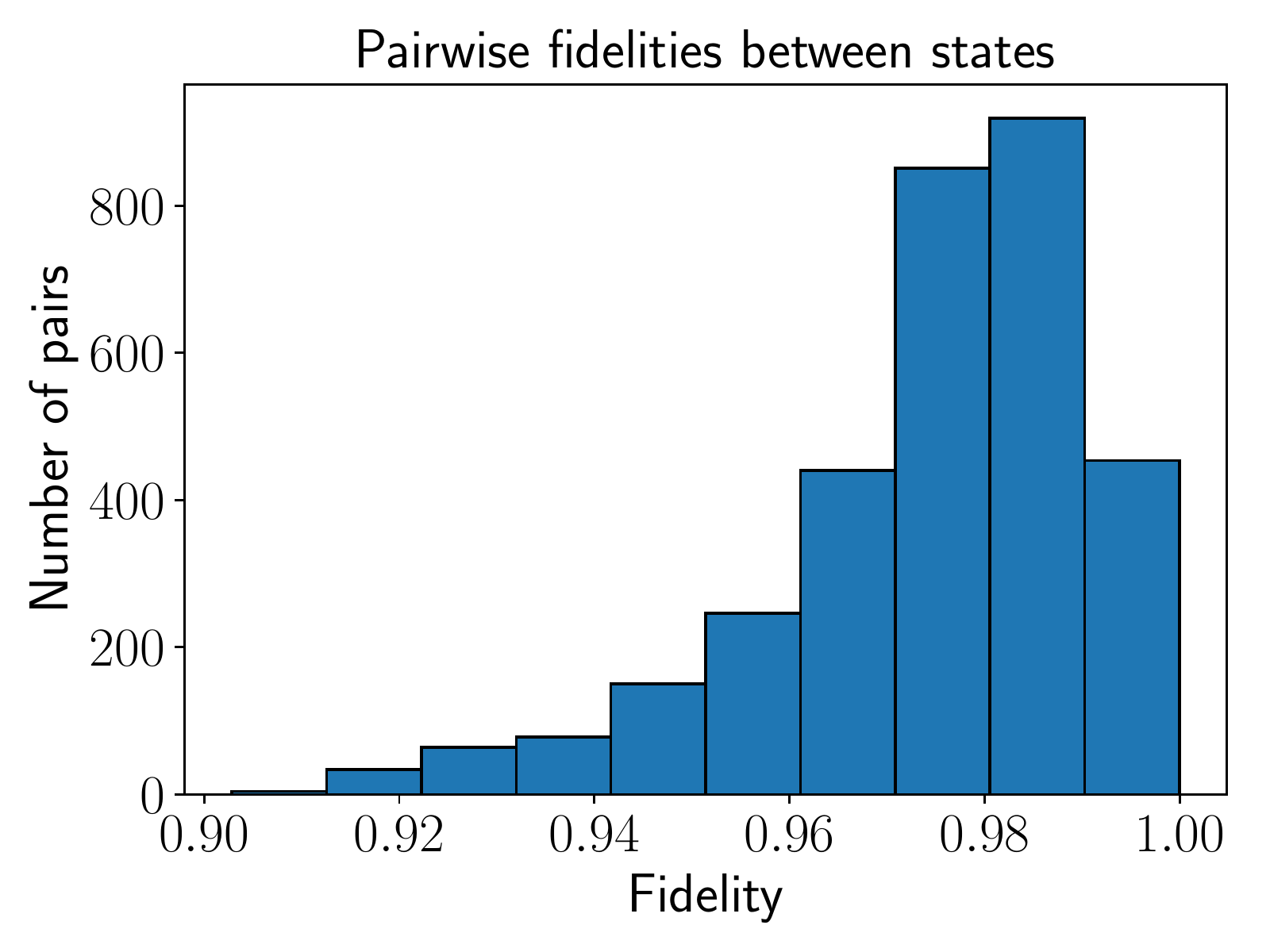}
    \caption{Pairwise fidelities between reconstructed states from 80 simulation runs with the same parameters; 500 trajectories and 20 tomography angles, with $\Omega=1/\sqrt{8}$ and $T=4$.}
    \label{fig:pair_fidelity}
\end{figure}

The distribution of pairwise fidelities for the 80 simulations with 500 trajectories and 20 angles can be seen in Fig.~\ref{fig:pair_fidelity}. The pairs are composed of all 3160 combinations of the 80 states. 
 
\begin{figure}[hbt!]
    \centering
    % exjobb_program/sensitivity/boxplot_negativity.py
    \includegraphics[width=\columnwidth]{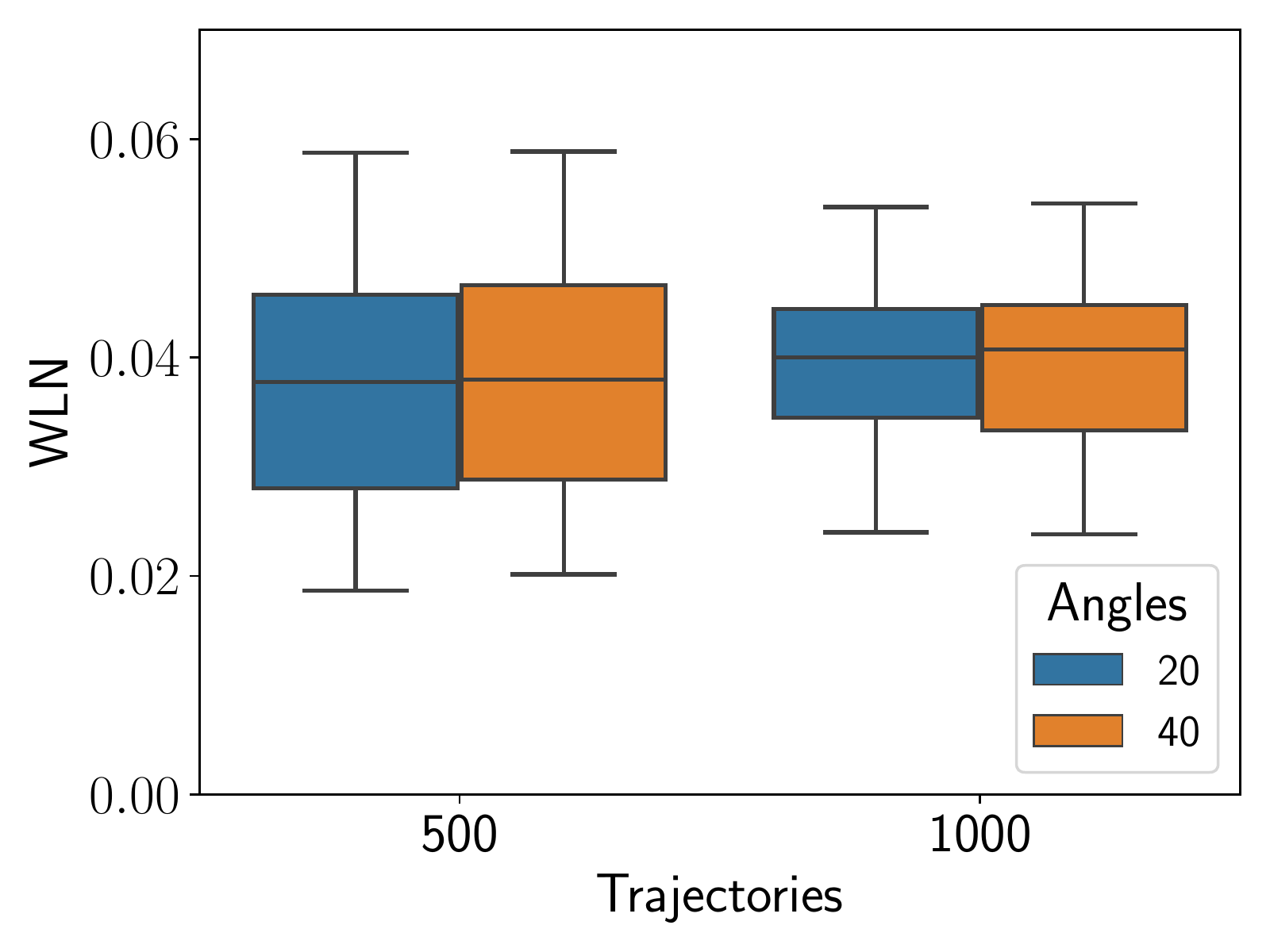}
    \caption{Wigner logarithmic negativity for 80 reconstructions from the atom in the semi-infinite waveguide at the incoherent point.}
    \label{fig:box_neg}
\end{figure}

\begin{figure}[hbt!]
% exjobb_program/sensitivity/analysis.py
    \centering
    \subfloat[Variations in the purity of 80 reconstructed states. The box shows the quartiles of the dataset while the whiskers extend to show the rest of the distribution. The horizontal line indicates the median.]{
        \includegraphics[width=\columnwidth]{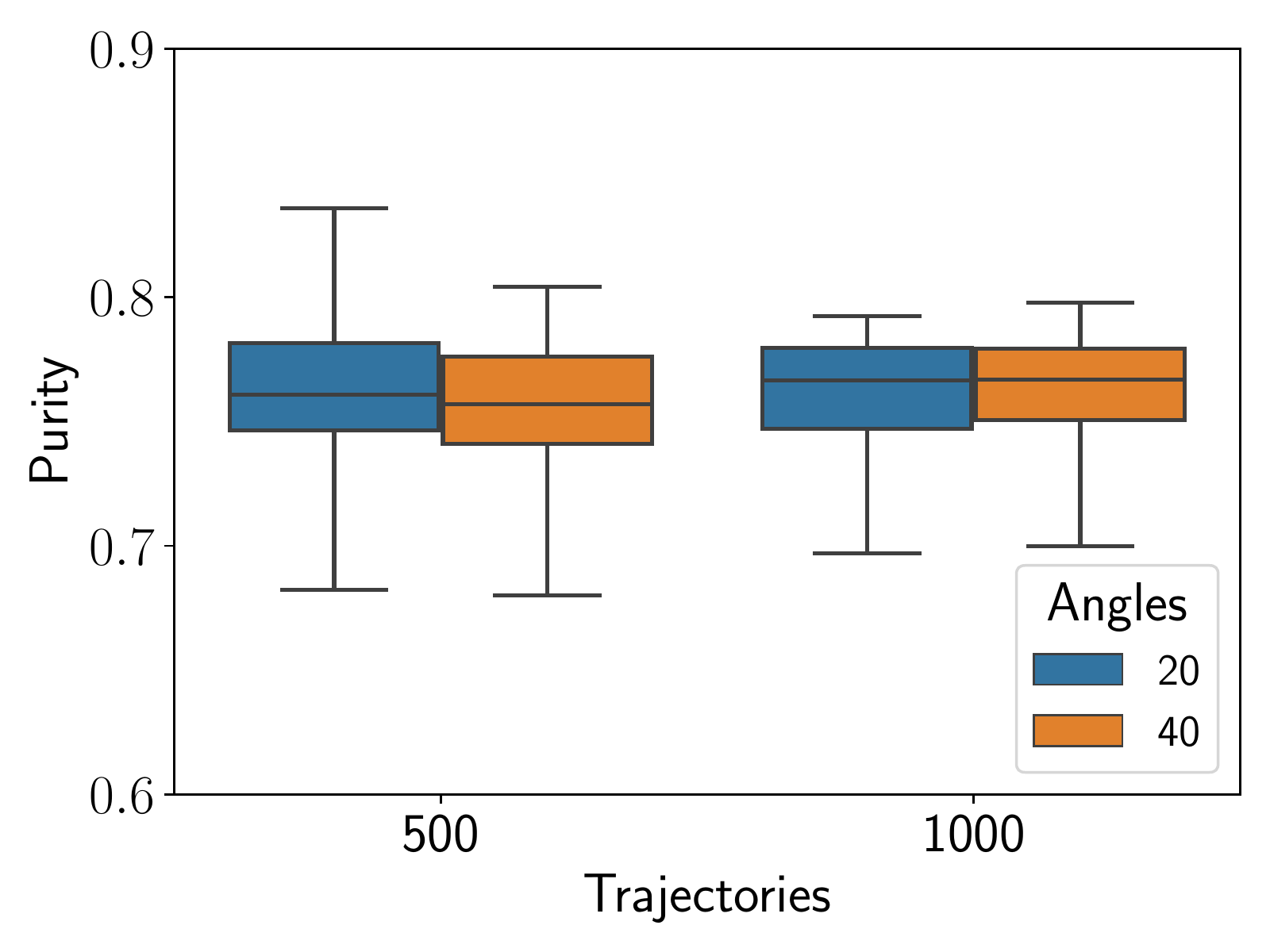}
        \label{fig:box_pur}
        }

    \subfloat[Scatterplot of the Wigner logarithmic negativity and the purity of the state, for 1000 trajectories and 20 angles. The Pearson correlation coefficient is $-0.2$.]{
        % exjobb_program/sensitivity/ corr_purity_negativity.py
        \includegraphics[width=\columnwidth]{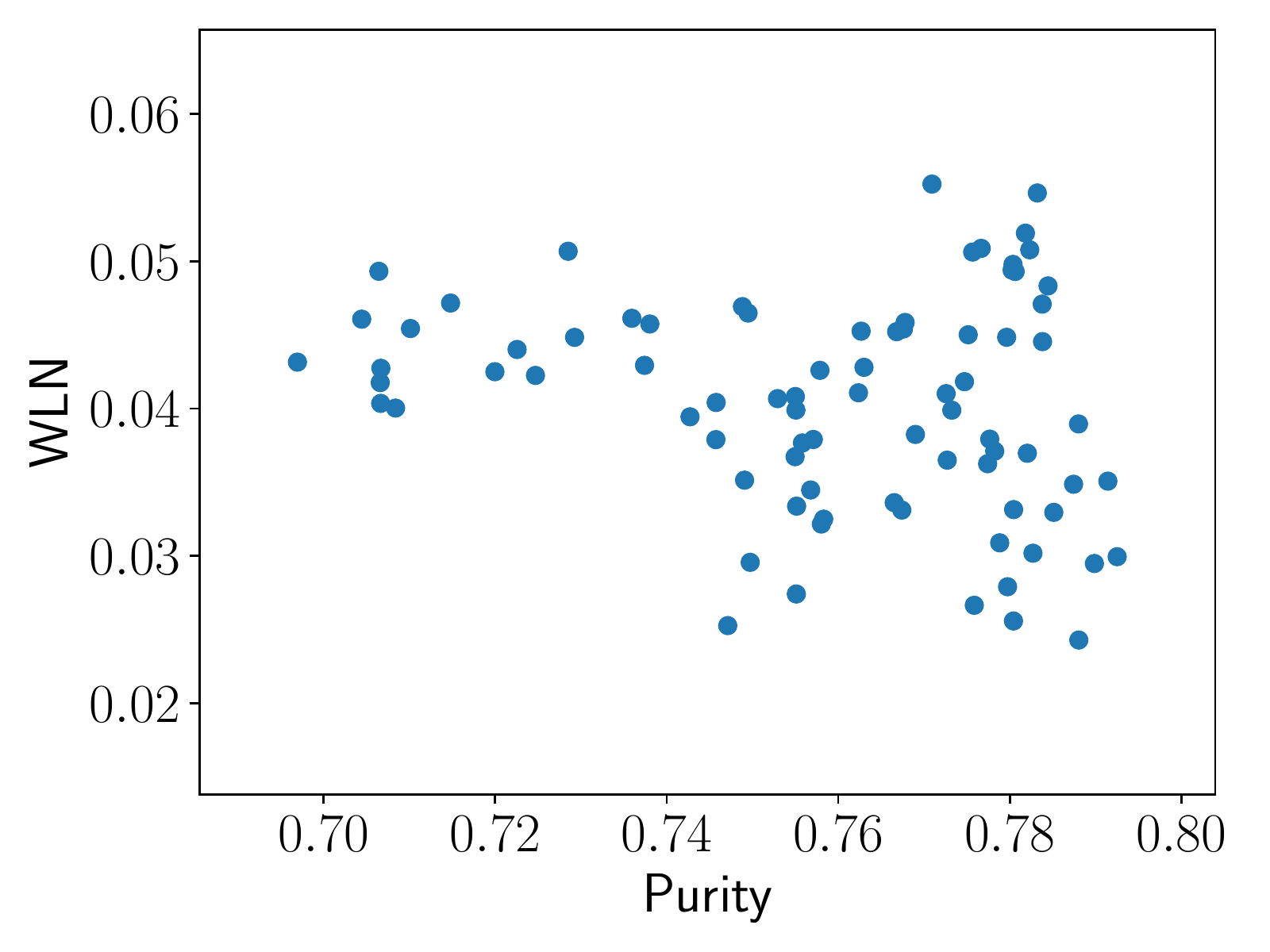}
        \label{fig:corr_pur}
        }
    \caption{Boxplot showing the variation in the purity of reconstructed states, and a scatterplot that shows no clear correlation between the purity and the WLN.}
    \label{fig:box_and_corr_pur}
\end{figure}

\begin{figure}[hbt!]
% exjobb_program/sensitivity/analysis.py
    \centering
    \subfloat[Variations in the single-photon population of 80 states reconstructed with identical parameters.]{
        \includegraphics[width=\columnwidth]{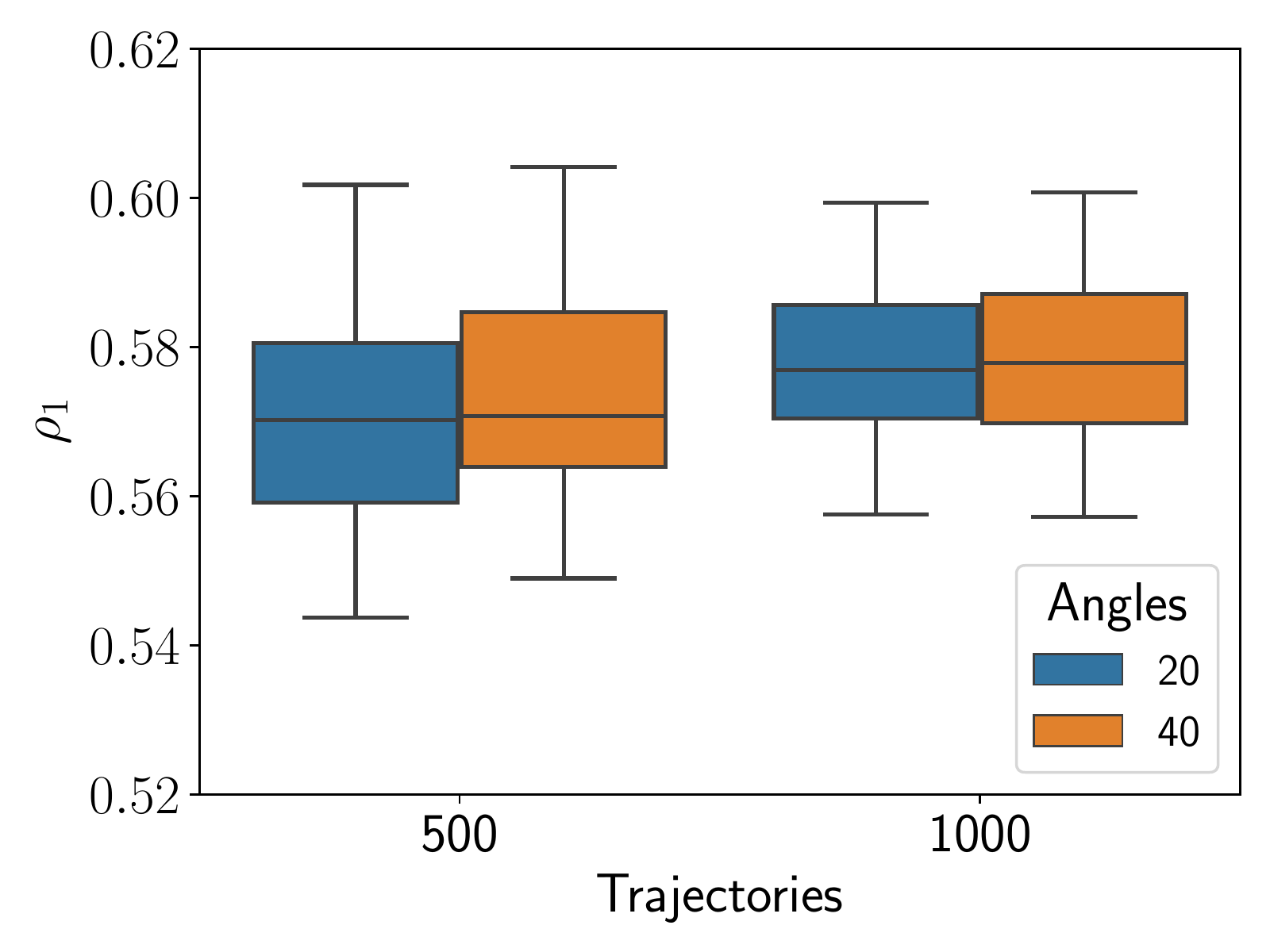}
        \label{fig:box_pop}
        }
    
    \subfloat[Scatterplot of the Wigner logarithmic negativity and the single-photon population, for 1000 trajectories and 20 angles. There is a clear correlation. and the Pearson correlation coefficient is $0.85$.]{
    %exjobb_program/sensitivity/corr_1photon_negativity.py
        \includegraphics[width=\columnwidth]{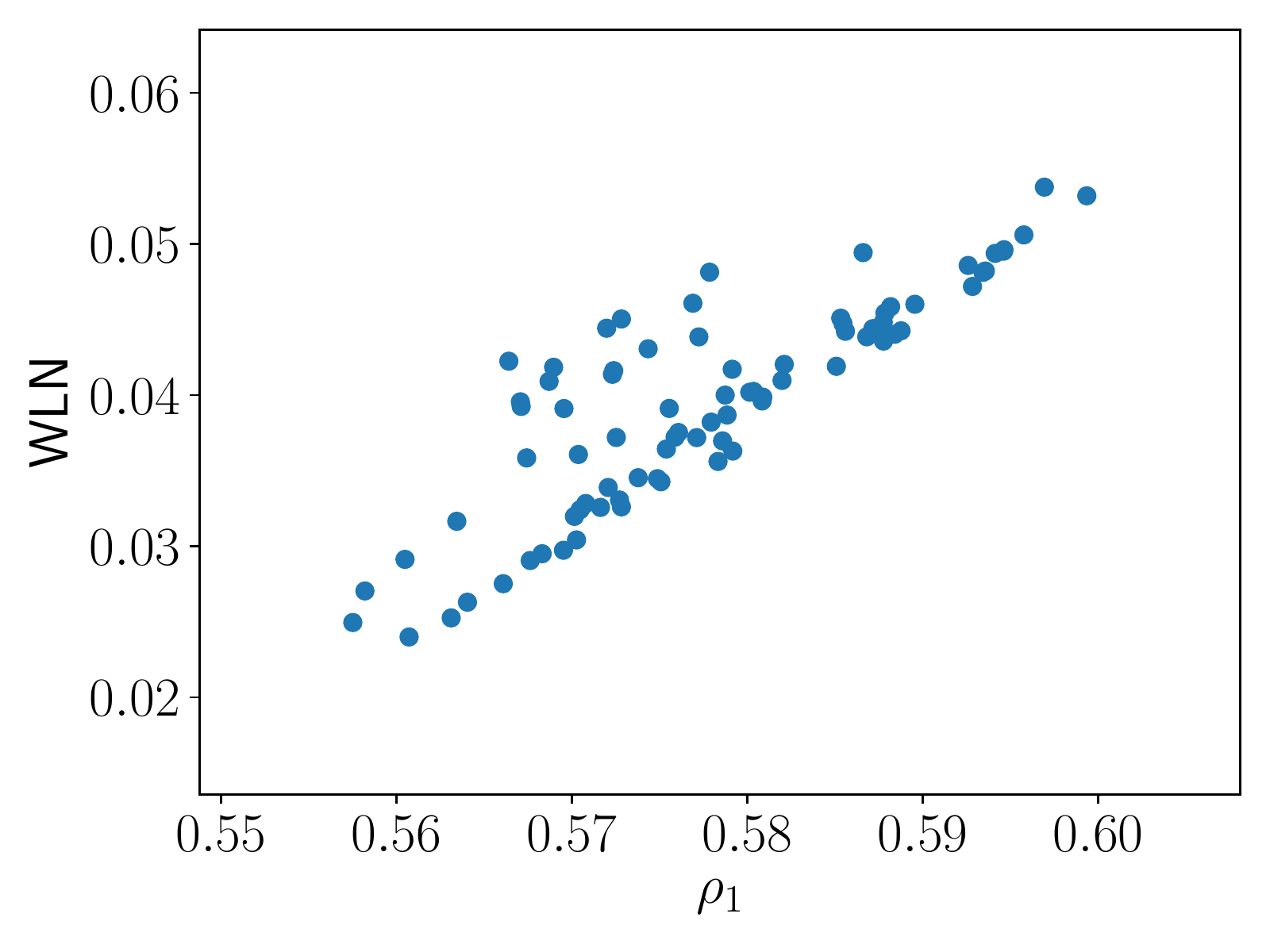}
        \label{fig:corr_pop}
        }
    \caption{Boxplot of the variation in single-photon content $\rho_1$, and a scatterplot that displays a linear correlation between $\rho_1$ and the WLN.}
    \label{fig:box_and_corr_pop}
\end{figure}

The data of WLN for Figs.~\ref{fig:only_dephasing},~\ref{fig:rad_and_dephasing} and~\ref{purity} are averages of the results from the 80 reconstructions. A single dataset only produced a very noisy curve, and even with averaging there are irregularities. The variation of the WLN is displayed in Fig.~\ref{fig:box_neg}. Since the mean WLN is very small, the relative variance is large. 
In order to find what creates the variation in the WLN we look at the properties that have been established in Section~\ref{results} to influence it: purity and single-photon population of the state. Figure~\ref{fig:box_pur} shows a boxplot of the distribution of purities, and Fig.~\ref{fig:corr_pur} shows a scatterplot containing the purity of the state and the corresponding WNL. There is visually no clear correlation. The variation in the single-photon population $\rho_1$ is seen in Fig.~\ref{fig:box_pop}. There is an obvious correlation between $\rho_1$ and the WLN, shown in the scatterplot~\ref{fig:corr_pop}, which is consistent with the results in previous sections.

%\subsection{statistics}

\FloatBarrier

%\afterpage{\clearpage}

\bibliography{ref}

\end{document}